\documentclass[twocolumn,secnumarabic,amssymb, nobibnotes, aps, pra, showkeys]{revtex4-2}
\usepackage[utf8]{inputenc}
\usepackage{amsmath, physics, graphicx, appendix, siunitx, verbatim}
\usepackage[version=4]{mhchem}
\usepackage{nicefrac}
\newcommand{\PMA}{Division of Physics, Math and Astronomy, California Institute of Technology, Pasadena, California 91125, USA}

\DeclareMathOperator{\Var}{Var}

\begin{document}

\title{Quantum Control of Two Critically Dressed Spin $\nicefrac{1}{2}$ Species in Magnetic Fluctuations}

\author{Raymond Tat}
\affiliation{\PMA}
\author{C. M. Swank}
\affiliation{\PMA}
\date{\today}


\begin{abstract}
The neutron electric dipole moment experiment at the Spallation Neutron Source (nEDM@SNS experiment) proposes to measure the nEDM using the spin-dependent capture cross section of neutrons on $^3$He. The critical dressing mode of this experiment uses an oscillating magnetic field to dress the gyromagnetic ratios of neutrons and $^3$He to the same value. While this technique grants increased sensitivity to the nEDM by improving the signal-to-noise ratio, this mode of measurement also introduces additional noise from the power supply used to drive the dressing field. This can lead to randomly fluctuating magnetic fields which cause the spins of neutrons and $^3$He to drift apart over time. Here we use second-order time-dependent perturbation theory to compute relaxation and frequency shifts due to fluctuations in the dressing field in terms of the magnetic field noise power spectrum and compare these calculations to numerical solutions obtained by integrating the Bloch equations. We then use these results to develop mitigation strategies for this type of noise. Furthermore, we report on spin dressing modulation techniques that significantly amplify coherence times for the critically dressed system, and attempt to quantify the coherence time achievable . 
\end{abstract}

\maketitle
%\tableofcontents

\section{Introduction}
When a particle precessing in a static magnetic field is exposed to an off-resonant oscillating orthogonal magnetic field, the effective gyromagnetic ratio of the particle - and thus its precession frequency - is modified. This phenomenon is known as spin dressing and was first proposed in reference \cite{cohen-tannoudji}. Spin dressing is found to have applications across multiple sub-fields in quantum information and fundamental physics, including increasing coherence time of free induction decay (FID) in relatively large magnetic field gradients \cite{BevilaqueGradientCancel,swank}. Multiple dressing frequencies tuned to extend the coherence of atomic clocks from field gradients was first reported in reference \cite{KazakovMagicDressing2015}, the authors of reference \cite{BoothMultiDressing2018} used this technique to mitigate Stark shifts. With multiple dressing field directions, enhanced spin manipulation for quantum information is achievable, and can even accelerate the effective Larmor precession; this is discussed in references \cite{Bevilacqua2020,bevilacqua2021harmonic}.  In reference \cite{golub1994}, the authors found that critical spin dressing (CSD), the simultaneous dressing of two spin species to the same Larmor frequency, can be applied for a very sensitive measurement of the neutron electric dipole moment (nEDM). A detailed investigation in reference \cite{apparatus} found that statistical uncertainty is doubled from this technique. With modulated CSD, described in references \cite{golub1994,swank}, systematic sensitivity from phase fluctuations not associated with noise in the spin dressing field, e.g. external magnetic field drifts, is significantly reduced without needing any correction from a comagnetometer. For example, with modulation at the angular frequency $\omega_m$ phase accumulation from holding field drifts in the apparatus will be modified by a factor $\sim1/\sqrt{\omega_m}$.

In the nEDM@SNS experiment ultracold neutrons (UCN) are confined to a measurement cell with dimensions $L_x \times L_y \times L_z = \SI{40 x 10.2 x 7.6}{\cm}$ in solution with polarized $^3$He in superfluid $^4$He. We detect the spin-dependent capture rate of neutrons on $^3$He through the reaction $\mathrm{n}+{^3\mathrm{He}}\rightarrow{^3\mathrm{H}}+p+764~ \mathrm{keV}$. The critical dressing mode of the experiment uses a strong oscillating magnetic field to dress gyromagnetic ratios of neutrons and \ce{^3He} to the same value. This improves sensitivity to the nEDM by allowing continuous operation at the most sensitive relative phase between the neutron and \ce{^3He} \cite{apparatus}
for a given \ce{^3He} concentration. However, this mode of measurement introduces additional noise, as current fluctuations in the spin dressing coil will lead to fluctuating magnetic fields. These fluctuating magnetic fields can cause the spins of neutrons and \ce{^3He} to drift apart over time. 

Here we use second-order time-dependent perturbation theory to compute relaxation and frequency shifts due to dressing field noise in terms of the dressing field noise power spectrum. We similarly derive run-to-run variance of the capture rate signal from dressing field fluctuations arising from current noise in a dressing coil, and compare them to simulations. We discuss the absence of run-to-run signal variance due to magnetic field gradients in either the static holding field or the dressing field. Finally, we propose methods to extend the coherence of the critically dressed system and mitigate the effects of field fluctuations caused by either current fluctuations in the dressing coil or from magnetic field gradients.

\section{Spin Dressing Hamiltonian}
A spin-$\nicefrac{1}{2}$ system with gyromagnetic ratio $\gamma$ precessing in a constant holding field $B_0 \hat{z}$ and a strong oscillating magnetic dressing field $B_1 \cos(\omega t) \hat{x}$ is described by the Hamiltonian
\begin{equation}
    H = \omega a^\dagger a + \frac{\Omega}{2} \sigma_x (a + a^\dagger) + \frac{\omega_0}{2} \sigma_z,
\end{equation}
where $\sigma_x$, $\sigma_y$, and $\sigma_z$ are the usual Pauli spin matrices, $a$ is the lowering operator for the photon field, $\omega_0=\gamma B_0$ is the undressed frequency of a spin in the holding field, defined along $\hat{z}$, and $\Omega$ is a coupling constant given by
\begin{equation}
    \Omega = \frac{\gamma B_1}{2 \lambda^{1/2}},
\end{equation}
where $\lambda = \expval{n}$ is the average photon number. We are interested in how applying a small fluctuating magnetic field $\delta B(t)$ with a known power spectrum would affect the time evolution of an initial state $\ket{\psi_0}$ under this Hamiltonian.

\section{Theory}
\subsection{Time-Dependent Perturbation Theory} \label{sec:tdpt}
Suppose that at time $t=0$, a system is in the state $\ket{\psi_0}$, and we wish to compute the expectation value of some operator $A_0$ at some future time $t$. Given a time-dependent perturbation $V(t)$ to a Hamiltonian $H$, The time-evolution operator $T(t)$ to second order in time-dependent perturbation theory is given by
\begin{align}
\begin{split}
    T(t) &= 1 - i\int_0^t dt' V_I(t') - \int_0^t \int_0^{t'} dt' dt'' V_I(t') V_I(t'') \\
    &+ O(V^3),
\end{split}
\end{align}

where $V_I(t) \equiv e^{iHt}V(t)e^{-iHt}$ is the perturbation evaluated in the interaction picture.
Thus, we need to compute $\bra{\psi(t)} A_0 \ket{\psi(t)} = \bra{\psi_0} T(t)^\dagger A T(t) \ket{\psi_0}$, where $A$ is likewise evaluated in the interaction picture, i.e. $A(t) \equiv e^{iHt}A_0e^{-iHt}$.

If $V(t)$ is proportional to $\delta B(t)$, then we can decompose $V_I(t)$ as a sum of complex exponentials as follows:
\begin{equation}
    V_I(t) = \left( \sum_{j} Q_j e^{-i\omega_j t} + Q_j^\dagger e^{i\omega_j t}\right) \delta B(t),
\end{equation}
where $Q_j$ are time-independent operators. Provided that $\delta B(t)$ has zero mean, the terms of $T^\dagger A T$ which are linear in $\delta B(t)$ will vanish when we compute an expectation value over functions $\delta B(t)$. Therefore we need only consider the quadratic terms, which can be decomposed as
\begin{equation}
    \sum_{j, k} f(Q_j, Q_k) \int_0^t dt' \int_0^{t'} dt'' e^{-i \omega_j t'} e^{i \omega_k t''} \delta B(t') \delta B(t''),
\end{equation}
where $f(Q_j, Q_k)$ is some operator which is a function of only $Q_j$, $Q_k$, and their Hermitian conjugates. We now evaluate the expectation value of the integral over functions $\delta B(t)$.
\begin{widetext}
\begin{align}
   &\int_0^t dt' \int_0^{t'} dt'' e^{-i \omega_j t'} e^{i \omega_k t''} \expval{\delta B(t') \delta B(t'') }, \\
   &= \int_0^t dt' \int_0^{t'} dt'' e^{-i (\omega_j - \omega_k) (t' + t'')/2} e^{-i (\omega_j + \omega_k) (t' - t'')/2} \expval{\delta B(t') \delta B(t'') }, \\
   &= \frac{1}{2}\left( \int_0^{2t} d\bar{t} e^{-i \Delta \omega \bar{t}} \right) \left(\int_0^{t - \abs{t - \bar{t}}} d\Delta t e^{-i \bar{\omega} \Delta t} \expval{\delta B\left(\frac{\Delta t}{2} + \bar{t}\right) \delta B\left(-\frac{\Delta t}{2} + \bar{t}\right) }\right) \label{eq:variable sub},\\
   &\approx \frac{1}{2}\left( \int_0^{2t} d\bar{t} e^{-i \Delta \omega \bar{t}} \right) \left(\int_0^{\infty} d\Delta t e^{-i \bar{\omega} \Delta t} R_{\delta B}(\Delta t)\right),
   \label{eq:factor frequency}
\end{align}
\end{widetext}
where in equation \ref{eq:variable sub} we have made the variable substitutions
\begin{align}
    \bar{\omega} &\equiv \frac{\omega_1 + \omega_2}{2}, \\
    \Delta \omega &\equiv \omega_1 - \omega_2, \\
    \bar{t} &= \frac{t' + t''}{2} \quad \text{(integration variable)}, \\
    \Delta t &= t' - t'' \quad \text{(integration variable)} ,\\
    R_{\delta B}(\Delta t) &= \expval{\delta B(\Delta t) \delta B(0) }.
\end{align}
In equation \ref{eq:factor frequency}, we assume that $\delta B(t)$ is stationary and has a short correlation time compared to $t$, and so it is valid to replace the limit of the second integral with infinity. The first integral in equation \ref{eq:factor frequency} only gros with with time if $\Delta \omega \neq 0$, and therefore any term where $\Delta \omega \neq 0$ can be neglected. From this we conclude that for the purpose of noise analysis, it is sufficient to consider the individual frequency components of $V_I(t)$ independently. Therefore without loss of generality, we can write
\begin{equation}
    V_I(t) = (Qe^{-i \omega t} + Q^\dagger e^{i \omega t}) \delta B(t),
\end{equation}
$T^\dagger A T$ can now be expressed in terms of $Q$ and $A$ to second order in $V_I$:
\begin{align}
\begin{split}
    T^\dagger A T ={}& A + \int_0^t \int_0^t dt' dt'' V_I(t'')AV_I(t') \\
     -& \int_0^t dt' \int_0^{t'} dt'' V_I(t'')V_I(t')A + AV_I(t')V_I(t''),
\end{split} \\
\begin{split} 
    ={}& A + uu^* \left( QAQ^\dagger + Q^\dagger A Q \right) - v (Q^\dagger Q A + A Q Q^\dagger)\\ &- v^* (Q Q^\dagger A + A Q^\dagger Q),
\end{split}\label{eq:tat expansion}
\end{align}
where we have defined the integrals
\begin{align}
    u(t; \omega) &\equiv \int_0^t dt' e^{-i \omega t'} \delta B(t'), \\
    v(t; \omega) &\equiv \int_0^t dt' \int_0^{t'} dt'' e^{-i \omega (t' - t'')} \delta B(t') \delta B(t'').
\end{align}
We can further simplify equation \ref{eq:tat expansion} by noting that
\begin{align}
    v + v^* &= \int_0^t dt' \int_0^{t'} dt'' \left(e^{-i \omega \Delta t} + e^{+i \omega \Delta t}\right) \delta B(t') \delta B(t'') ,\\
    &= \int_0^t \int_0^{t} dt' dt'' e^{-i \omega \Delta t} \delta B(t') \delta B(t''), \\
    &= u u^* .\label{eq:vvuu}
\end{align}
If we now collect the terms proportional to $v$ and $v^*$, we get
\begin{equation}
T^\dagger A T = A + [Q^\dagger, [A, Q]] v  + [Q, [A, Q^\dagger]] v^*.
\end{equation}
We can alternatively choose to collect terms proportional to $\Re(v)$ and $\Im(v)$, in which case we get
\begin{equation}
\begin{split}
T^\dagger A T ={}& A + \left([Q, [A, Q^\dagger]] + [Q^\dagger, [A, Q]]\right) \Re(v) \\
&+ i [A, [Q^\dagger, Q]] \Im(v). \label{eq:expectation shortcut}
\end{split}
\end{equation}
This allows us to calculate the expectation value of an operator $A$ at time $t$ in the interaction picture. In other cases it may also be useful to calculate the variance of $A$. In the nEDM@SNS experiment for example, variance in the capture rate signal, which depends linearly on the dot product between neutron and \ce{^3He} spins, would lead to uncertainties in the nEDM measurement. In this particular case we are concerned not with the quantum mechanical variance of $\vec{\sigma_1} \cdot \vec{\sigma_2}$ (which is nonzero even if the magnetic field has no fluctuations), but rather in how the magnetic field perturbs the spin vectors classically. In other words, we wish to find the variance of the quantum mechanical expectation value of $\vec{\sigma_1} \cdot \vec{\sigma_2}$. We thus define the classical variance of an operator $A$ as
\begin{equation}
    \Var_{cl}(A) \equiv \expval{\abs{\expval{A}}^2}_{\delta B} - \abs{\expval{\expval{A}}_{\delta B}}^2,
\end{equation}
where $\expval{\cdot}_{\delta B}$ denotes an average over random functions $\delta B(t)$, while $\expval{\cdot}$ denotes a quantum mechanical expectation value, i.e. $\expval{A} \equiv \bra{\psi} A \ket{\psi}$. A similar analysis as in (\ref{eq:tat expansion}) yields, in the case that $A$ is Hermitian,
\begin{equation}
    \Var_{cl}(T^\dagger A T) = 4\expval{[Q^\dagger, A]}\expval{[A, Q]} \Re(v). \label{eq:variance}
\end{equation} 
These results can also be applied to a pair of non-interacting spin-1/2 systems through use of the tensor product. Suppose that in the interaction picture the time-dependent perturbations for each of the two spins are given by
\begin{align}
V_1(t) &= (Q_1e^{-i \omega t} + Q_1^\dagger e^{i \omega t}) \delta B(t), \\
V_2(t) &= (Q_2e^{-i \omega t} + Q_2^\dagger e^{i \omega t}) \delta B(t).
\end{align}
If we wish to compute the expectation value $\expval{A}$ from an initial state $\ket{\psi}_0 = \ket{\psi_1}_0 \otimes \ket{\psi_2}_0$, then
we can apply equation \ref{eq:expectation shortcut} or \ref{eq:variance} by substituting
\begin{equation}
    Q \rightarrow Q_1 \otimes I_2 + I_1 \otimes Q_2,
\end{equation}
where $I_1$ and $I_2$ are identity operators.

Lastly, we express $v$ in terms of the power spectrum $S(\omega)$ of $\delta B(t)$. The power spectrum is defined as
\begin{equation}
    S(\omega) \equiv \int_{-\infty}^{\infty} dt e^{-i\omega t} R_{\delta B}(t).
\end{equation}
From equation \ref{eq:factor frequency}, we have
\begin{align}
     v(t; \omega) &= \frac{1}{2}\left( \int_0^{2t} d\bar{t} \right) \left(\int_0^{\infty} d\tau e^{-i \omega \tau} R_{\delta B}(\tau)\right), \\
     &= \frac{t}{2 \pi} \int_0^\infty d\tau e^{-i \bar{\omega} \tau} \int_{-\infty}^{\infty} d\omega' S(\omega') e^{i \omega' \tau}, \\
     &= \frac{t}{2 \pi} \int_{-\infty}^{\infty} d\omega' S(\omega') \int_0^\infty d\tau e^{i (\omega' - \omega) \tau}, \\
     &= \frac{t}{2 \pi} \int_{-\infty}^{\infty} d\omega' S(\omega') \left( \pi \delta (\omega' - \omega) - \frac{i}{\omega' - \omega} \right), \\
     &= t \left (\frac{1}{2} S(\omega) - \frac{i}{2 \pi} \int_{-\infty}^{\infty} d\omega' \frac{S(\omega')}{\omega' - \omega} \right).
\end{align}

\subsection{Initial State}
An oscillating magnetic field corresponds quantum mechanically to a coherent state, denoted by $\ket{\alpha}$. A coherent state is an eigenstate of the lowering operator with eigenvalue $\alpha$. The average photon number in such a state is given by $\abs{\alpha}^2 = \lambda$, and so we separate $\alpha$ into its magnitude and complex phase as $\alpha \equiv \sqrt{\lambda} e^{i \theta}$. In the appendix it is shown that a complex phase of $\alpha$ is equivalent to a rotation of the neutron and $^3$He spins. The initial state in the following is taken to be
\begin{equation}
    \ket{\psi_0} = \ket{\alpha} \ket{s},
\end{equation}
where $\ket{s}$ is the initial spin state, or if we are considering both the neutron and $^3$He spins,
\begin{equation}
    \ket{\psi_0} = \ket{\alpha} \ket{s_1} \ket{s_2}.
\end{equation}

\subsection{Computation of Matrix Elements}
We now compute the $Q$ operators (and thus $V_I(t)$) for the spin dressing Hamiltonian in the case where the time-dependent fluctuating magnetic field is parallel to $\hat{x}$. This calculation will proceed in two steps. We first apply a carefully chosen unitary transform to the Hamiltonian in order to compute its eigenstates, and then compute the matrix elements of the operator $\sigma_x$ with respect to these states. We then transform these matrix elements into the interaction picture by multiplying each matrix element by the appropriate time-dependent phase factor. In the limit where $\lambda \gg 1$ and $\omega \gg \omega_0$, the eigenstates and eigenvalues of the spin dressing Hamiltonian can be found approximately by expressing $H$ as $H = H_0 + H_z$, where
\begin{align}
    H_0 &= \omega a^\dagger a + \frac{\Omega}{2} \sigma_x (a + a^\dagger), \\
    H_z &= \frac{\omega_0}{2} \sigma_z.
\end{align}
As shown in \cite{cohen-tannoudji}, $H_0$ can be diagonalized by considering each of the two eigensubspaces of $\sigma_x$ individually. The Hamiltonians of the two subspaces are given by
\begin{equation}
H_\epsilon = \omega a^\dagger a + \frac{\Omega}{2} \epsilon (a + a^\dagger), \\
\end{equation}
where $\epsilon \in \{1, -1\}$ corresponds to the eigenvalues of $\sigma_x$. The part of the Hamiltonian contained in $H_\epsilon$ is diagonalized by applying the displacement operator $D(\epsilon \Omega/2\omega)$, which is defined by its action on the raising and lower operators:
\begin{align}
    D^\dagger(\eta) a D(\eta) &\equiv a + \eta,  \\
    D^\dagger(\eta) &= D(-\eta).
\end{align}
Explicitly, the displacement operator is given by
\begin{equation}
    D(\eta) \equiv e^{\eta a^\dagger - \eta^* a}.
\end{equation}
Note that while the definition of the displacement operator $D(\eta)$ allows for complex $\eta$, here we restrict our attention to the case where $\eta$ is real. Applying this operator to $H_\epsilon$ and abbreviating $D(\epsilon \Omega/2\omega)$ as $D_\epsilon$ yields
\begin{align}
    D_\epsilon H_\epsilon D_\epsilon^\dagger ={}& \omega D_\epsilon a^\dagger a D_\epsilon^\dagger + \frac{\epsilon \Omega}{2} D_\epsilon(a + a^\dagger) D_\epsilon^\dagger\\
    \begin{split}
    ={}& \omega \left(a^\dagger - \frac{\epsilon \Omega}{2\omega}\right)\left(a - \frac{\epsilon \Omega}{2\omega}\right) \\ &+ \frac{\epsilon\Omega}{2}\left(a + a^\dagger - \frac{\epsilon\Omega}{\omega}\right)
    \end{split} \\
    ={}& \omega a^\dagger a - \frac{\Omega^2}{4\omega}.
\end{align}
From $D_\epsilon$ we can construct a unitary operator $U$, which applies $D_\epsilon$ to the appropriate eigensubspace of $\sigma_x$. Explicitly,
\begin{equation}
    U \equiv D\left(\frac{\Omega}{2 \omega}\right) \ket{+_x}\bra{+_x} + D^\dagger\left(\frac{\Omega}{2 \omega}\right) \ket{-_x}\bra{-_x},
\end{equation}
where $\ket{\pm_x}$ are the eigenvectors of $\sigma_x$.
Applying $U$ to $H_0$ yields
\begin{equation}
    UH_0U^\dagger = \omega a^\dagger a - \frac{\Omega^2}{4\omega}.
\end{equation}
% \begin{widetext}
% \begin{align}
%     U H_0 U^\dagger ={}& \omega \left[D a^\dagger a D^\dagger \ket{+_x}\bra{+_x} + D^\dagger a^\dagger a D \ket{-_x}\bra{-_x} \right]
%     + \frac{\Omega}{2} \left[ D (a + a^\dagger) D^\dagger \ket{+_x}\bra{+_x} - D^\dagger (a + a^\dagger) D \ket{-_x}\bra{-_x}\right] ,\\
%     \begin{split}
%     ={}& \omega \left[ \left(a^\dagger - \frac{\Omega}{2\omega} \right) \left(a - \frac{\Omega}{2\omega} \right)\ket{+_x}\bra{+_x}  + \left(a^\dagger + \frac{\Omega}{2\omega} \right) \left(a + \frac{\Omega}{2\omega} \right)\ket{-_x}\bra{-_x} \right]  \\
%     &+ \frac{\Omega}{2} \left [\left(a + a^\dagger - \frac{\Omega}{\omega} \right)\ket{+_x}\bra{+_x}  - \left(a + a^\dagger + \frac{\Omega}{\omega} \right)\ket{-_x}\bra{-_x} \right], 
%     \end{split} \\
%     ={}& \omega a^\dagger a - \frac{\Omega^2}{4\omega}.
% \end{align}
% \end{widetext}
We thus see that $U$ diagonalizes $H_0$, and that the eigenstates of $UH_0U^\dagger$ are $\ket{n}\ket{\pm_z}$. Meanwhile, $H_z$ in this basis becomes
\begin{equation}
    U H_z U^\dagger = \frac{\omega_0}{2} \left[ D^\dagger\left(\frac{\Omega}{\omega}\right) \ket{+_x}\bra{-_x} + D\left(\frac{\Omega}{\omega}\right) \ket{-_x}\bra{+_x} \right].
\end{equation}
For large photon number ($\lambda \gg 1$), the displacement operator can be approximated in terms of Bessel functions. Reference \cite{cohen-tannoudji} gives the matrix elements of $D(\eta)$ as
\begin{equation}
    \bra{n} D(\eta) \ket{n-q} = \bra{n} e^{\eta(a^\dagger - a)} \ket{n-q} = J_q(2 \eta \sqrt{\lambda}),
\end{equation}
where $J_q$ is a Bessel function of the first kind with order $q$. From this we derive
\begin{equation}
    D(\eta) \approx \sum_{n, q} J_q(2 \eta \sqrt{\lambda}) \ket{n+q}\bra{n}. \label{eq:bessel-approx}
\end{equation}
With this approximation, the full Hamiltonian $H$ in the displaced basis is

\begin{equation}
\begin{split}
    U H U^\dagger ={}& \omega a^\dagger a \\
    &+ \frac{\omega_0}{2}J_0(x) \sigma_z \\
    &- \frac{\omega_0}{2}J_1(x) \left(\sum_n\ket{n}\bra{n+1} - \ket{n+1}\bra{n}\right) i \sigma_y \\
    &+ \frac{\omega_0}{2}J_2(x) \left(\sum_n\ket{n}\bra{n+2} + \ket{n+2}\bra{n}\right)\sigma_z \\
    &+ \mathcal{O}(q\geq3). 
\end{split}
\end{equation}
where we have defined the spin dressing parameter $x \equiv 2 \Omega \sqrt{\lambda} /\omega = \gamma B_1/\omega$. In reference \cite{swank}, the matrix elements of various perturbations are calculated in time-dependent perturbation theory. Here, we extend those results to the case where $\omega_0$ may not be small compared to $\omega$ by incorporating the first-order correction to the eigenstates of $UH_0U^\dagger$. We first consider the energy correction.  Let the eigenstates of $UHU^\dagger$ be denoted by $\widetilde{\ket{n, \pm_z}}$. Treating $H_z$ as a perturbation on $H_0$, the energies of these states can be calculated to first order in $\omega_0/\omega$ using perturbation theory, as in reference \cite{cohen-tannoudji}. The energy shift is simply 
\begin{align}
    E_{n,\pm} &= E^{(0)}_{n, \pm} + \bra{n}\bra{\pm_z} H_z \ket{\pm_z} \ket{n} \\
    &= n \omega \pm \frac{\omega_0}{2}J_0(x).
\end{align}
$H_z$ thus lifts the degeneracy of $\ket{n}\ket{\pm_z}$ and leads to an effective Larmor frequency $\omega_0' \equiv \omega_0 J_0(x)$. We similarly derive the eigenstates of $UHU^\dagger$ to first order in $\omega_0/\omega$:
\begin{equation}
\begin{split}
    \widetilde{\ket{n, \pm_z}} &= \ket{n}\ket{\pm_z} 
    \\&\pm \frac{J_1(x) \omega_0}{2 \omega} \left(\ket{n-1}\ket{\mp_z} + \ket{n+1}\ket{\mp_z}\right) \\
    &\pm \frac{J_2(x)\omega_0}{4\omega} (\ket{n-2}\ket{\pm_z} - \ket{n+2}\ket{\pm_z}) \\
    &+ \mathcal{O}(q\geq3).
\end{split}
\end{equation}
We can now calculate the matrix elements of various time-dependent perturbations with respect to these eigenstates. For dressing field noise generated from fluctuations in the current provided by the power supply, the only relevant operator is $\sigma_x$, provided  the linearly polarized dressing field is homogeneous. Fortunately, $\sigma_x$ is unchanged by $U$, as $U \sigma_x U^\dagger = \sigma_x$. We obtain
\begin{align}
\begin{split}
    \widetilde{\bra{n, \pm_z}} \sigma_x \widetilde{\ket{n', \pm_z}} &= \pm \frac{J_1(x) \omega_0}{\omega} \delta_{1, \abs{n-n'}} \\
    &+ \mathcal{O}(q\geq3),
\end{split} \\
\begin{split}
    \widetilde{\bra{n, \pm_z}} \sigma_x \widetilde{\ket{n', \mp_z}} &= \delta_{n, n'} \pm \frac{J_2(x)\omega_0}{2 \omega}\delta_{2, \abs{n-n'}}\\ &+ \mathcal{O}(q\geq3).
\end{split}
\end{align}
We have truncated terms containing Bessel functions of order 3 or greater. From these matrix elements, we can now evaluate the perturbation in the interaction picture by computing the individual frequency components of $e^{iU H U^\dagger t} \sigma_x e^{-iU H U^\dagger t}$ using
\begin{equation}
    e^{iU H U^\dagger t} = \sum_{n, s} \widetilde{\ket{n, s}} \widetilde{\bra{n, s}} e^{i \omega_{n, s} t},
\end{equation}
where $\hbar \omega_{n, s}$ is the energy of the state $\widetilde{\ket{n, s}}$. Expanding $e^{iU H U^\dagger t} \sigma_x e^{-iU H U^\dagger t}$, we get
\begin{equation}
\begin{split}
    e^{iU H U^\dagger t}& \sigma_x e^{-iU H U^\dagger t} =\\ &\sum_{n, n', s, s'} \widetilde{\ket{n, s}} \widetilde{\bra{n, s}} \sigma_x \widetilde{\ket{n', s'}} \widetilde{\bra{n', s'}} e^{i \Delta \omega t},
\end{split}
\end{equation}
where $\Delta \omega$ is the frequency difference between $\widetilde{\ket{n, s}}$ and $\widetilde{\ket{n', s'}}$. Here we only consider the leading-order term of each frequency component, so we make the approximation that $\widetilde{\ket{n, s}} \widetilde{\bra{n', s'}} \approx \ket{n}\ket{s} \bra{n'}\bra{s'}$. With this approximation, we can now write
\begin{equation}
    e^{iU H U^\dagger t} \sigma_x e^{-iU H U^\dagger t} = W + W^\dagger,
\end{equation}
where 
\begin{equation}
\begin{split}
    W &= -\sigma_- e^{-i\omega_0' t} \\
    +& \sigma_z \sum_{n} \ket{n-1}\bra{n} \frac{J_1(x) \omega_0}{\omega} e^{-i \omega t} \\
    -& \sigma_- \sum_{n} \ket{n-2}\bra{n} \frac{J_2(x) \omega_0}{2 \omega} \left(e^{-i(2w + \omega_0')t} - e^{-i(2w - \omega_0')t}\right) \\
    +& \mathcal{O}(q\geq3).
\end{split}
\end{equation} \label{eq:W expanded}
We can simplify this expression further. In the classical limit, we can make the approximation that
\begin{equation}
    a \approx \sqrt{\lambda}\sum_{n} \ket{n-1}\bra{n},
\end{equation}
provided that $\lambda$ is large. Thus when applying operators of the form $\sum_{n} \ket{n-q}\bra{n}$ to a coherent state, we get
\begin{align}
    \left(\sum_{n} \ket{n-q}\bra{n}\right) \ket{\alpha} &\approx \frac{a^q}{\sqrt{\lambda^q}} \ket{\alpha}, \\
    &= e^{i q \theta} \ket{\alpha}.
\end{align}
This allows us to replace the terms operating on the photon field in $W$ with simple phase factors. Thus,
\begin{equation}
\begin{split}
    W ={}& -\sigma_- e^{-i\omega_0' t} \\
    &+ \sigma_z e^{i \theta} \frac{J_1(x) \omega_0}{\omega} e^{-i \omega t} \\
    &- \sigma_- e^{2 i \theta} \frac{J_2(x) \omega_0}{2 \omega} (e^{-i(2w + \omega_0')t} - e^{-i(2w - \omega_0')t})\\
    &+ \mathcal{O}(q\geq3).
\end{split}
\end{equation} \label{eq:W simplified}
Each independent frequency term of $W$ corresponds to a $Q$ operator. In particular for a perturbation $\delta H(t) = \sigma_x \delta B(t)$, the corresponding $Q$ operators are
\begin{align}
    Q_{\omega_0'} &= - \sigma_- \label{eq:Q_first},\\
    Q_{\omega} &= \frac{J_1(x) \omega_0}{\omega} \sigma_z, \\
    Q_{2\omega + \omega_0'} &= -\frac{J_2(x) \omega_0}{2\omega} \sigma_- ,\\
    Q_{2\omega - \omega_0'} &= \frac{J_2(x) \omega_0}{2\omega} 
    \sigma_+ \label{eq:Q_last}.
\end{align}
In writing these operators we have omitted any phase factors, as in both equations \ref{eq:expectation shortcut} and \ref{eq:variance}, $Q$ is always paired with $Q^\dagger$, and so any phase factor will be eliminated.
\\
\subsection{Explicit Calculation for Magnetic Field Fluctuations}
Recalling that equation \ref{eq:expectation shortcut} allows us to calculate the expectation value of an operator at any time in the interaction picture, and combining equations \ref{eq:expectation shortcut} and \ref{eq:Q_first}-\ref{eq:Q_last}, we can now explicitly compute the effects of fluctuations in the amplitude and phase of the dressing field. As in \cite{swank}, we compute relaxation and frequency shifts by taking the real and imaginary parts of $\expval{\sigma_x + i\sigma_y}$ in the interaction picture (rotating frame) for a spin which starts in the $+\hat{x}$ direction at time $t=0$, i.e. $\ket{\psi_0} = (\ket{+_z} + \ket{-_z})/\sqrt{2}$. Given a perturbation $\delta H(t) = \gamma \delta B(t) \sigma_x /2$, we obtain
\begin{align}
\begin{split}
    \frac{1}{T_2} ={}& -\dv{\expval{\sigma_x}}{t} = \frac{\gamma^2}{4} S(\omega_0') \\
    &+ \left(\frac{\gamma J_1(x)\omega_0}{\omega} \right)^2 S(\omega)  \\
    &+ \frac{1}{4}\left(\frac{\gamma J_2(x)\omega_0}{2 \omega} \right)^2  \left[S(2\omega - \omega_0') + S(2\omega + \omega_0')\right] \\
    &+ \mathcal{O}(q\geq3),
\end{split} \\
\begin{split}
    \delta \omega ={}& \dv{\expval{\sigma_y}}{t} = -\frac{\gamma^2}{4\pi}  \int_{-\infty}^{\infty} d\omega' \frac{S(\omega')}{\omega' - \omega_0'} \\ &- \frac{1}{4 \pi}\left(\frac{\gamma J_2(x)\omega_0}{2 \omega} \right)^2 \int_{-\infty}^{\infty} d\omega' S(\omega') \\
    &\times \left( \frac{1}{\omega' - 2\omega + \omega_0'} + \frac{1}{\omega' - 2\omega - \omega_0'} \right)\\
    &+ \mathcal{O}(q\geq3).
\end{split}
\end{align}
 We note that in the specific case where the dressing field fluctuations arise due to current fluctuations in the dressing coil, ``transverse relaxation" should be interpreted as run-to-run phase variation (rather than as depolarization during each individual run), as all neutrons would experience the same time-dependent magnetic field. Similarly, we can compute the run-to-run signal variance due to a fluctuating magnetic field for the case where both neutrons and \ce{^3He} experience the same time-dependent fluctuation. The capture rate of neutrons on $^3$He depends linearly on $\vec{\sigma_1} \cdot \vec{\sigma_2}$, and so we would like to compute the expectation value of this operator. However, the expectation values computed from time-dependent perturbation theory are in the interaction picture, so to convert these values to those actually observed in the lab frame we must find the variance of $U_1 U_2 e^{iH t}  \sigma_1 \cdot \sigma_2 e^{-iHt} U_2^\dagger U_1^\dagger$. Generically this is a time-dependent operator having terms which oscillate at frequencies that are linear combinations of the dressing field frequency $\omega$ and the dressed Larmor precession frequency $\omega_0'$. However, because capture signal rates in the nEDM@SNS experiment are not sufficiently high to resolve oscillations at these frequencies, we consider only the time-independent component of this operator. We can now calculate the variance in the capture rate due to a fluctuating magnetic field. Let $\vec{b_1}=\expval{\vec{\sigma_1}}(t=0)$ and $\vec{b_2}=\expval{\vec{\sigma_2}}(t=0)$ be the Bloch vectors of the neutron and $\ce{^3He}$ atom at time $t=0$. Then
\begin{widetext}
\begin{equation}
\begin{split}
\Var_{cl}(\sigma_1 \cdot \sigma_2)(t) ={}& \frac{1}{2} (\gamma_1 - \gamma_2)^2 \abs{\hat{z} \times (\vec{b_1} \times \vec{b_2})}^2 S(\omega_0') t \\
&+ 2 \left(\frac{\gamma_1 J_1(x_1)\omega_1 - \gamma_2 J_1(x_2)\omega_2}{\omega} \right)^2 \abs{\hat{z} \cdot (\vec{b_1} \times \vec{b_2})}^2 S(\omega) t \\
&+ \frac{1}{2} \left(\frac{\gamma_1 J_2(x_1)\omega_1 - \gamma_2 J_2(x_2)\omega_2}{2 \omega} \right)^2 \abs{\hat{z} \times (\vec{b_1} \times \vec{b_2})}^2 \left[S(2\omega - \omega_0') + S(2\omega + \omega_0')\right] t \\
&+ \mathcal{O}(q\geq3). \label{eq:QMvariance}
\end{split}
\end{equation}
\end{widetext}
The first term in equation (\ref{eq:QMvariance}) represents the variance due to noise at the effective Larmor frequency. However, if both spins start in the plane of precession this term will not contribute. The second term represents variance due to noise at the dressing frequency, which unlike the first term is maximized if the spins start in the plane of precession. The third term, like the first, is minimized when the spins start in the plane of the precession and is generated from noise at twice the dressing frequency shifted by the effective Larmor precession frequency. 

Equation (\ref{eq:QMvariance}) may appear to suggest that if $\vec{b_1}=\vec{b_2}$, one could altogether eliminate variance due to current fluctuations. However, under critical spin dressing the quantity proportional to the nEDM is the relative phase between neutron and \ce{^3He} spins, i.e. $\phi_2 - \phi_1 \propto d_n$. Meanwhile, the signal rate of nEDM@SNS is linear in $\sigma_1 \cdot \sigma_2$, which is in turn related to the relative phase by
\begin{align}
    \sigma_1\cdot\sigma_2=\cos(\phi_2-\phi_1),
\end{align}
provided both spins start in the plane of precession. Thus setting $\phi_2 = \phi_1$ would decrease sensitivity to an nEDM, as $\sigma_1\cdot\sigma_2$ would no longer be sensitive to the relative phase in this configuration. Furthermore, the variance of $\phi_2 - \phi_1$ is independent of the starting phase of either spin. Again assuming both spin start in the plane of precession,
\begin{align}
\hat{z} \cdot (\vec{b_1} \times \vec{b_2})=\sin(\phi_2-\phi_1),
\end{align}
The variance can be propagated using the Taylor expansion. We have,
\begin{align}
\Var_{cl}(\sigma_1 \cdot \sigma_2)\approx\abs{\sin(\phi_2-\phi_1)}^2\Var_{cl}(\phi_2-\phi_1).
\end{align}
Therefore, in the plane of precession the phase of the spin of one species with respect to the spin other the species does not matter, and the variance in relative phase will accumulate according to
\begin{align}
\Var_{cl}(\phi_2-\phi_1)\approx 2 \left(\frac{\gamma_1 J_1(x_1)\omega_1 - \gamma_2 J_1(x_2)\omega_2}{\omega} \right)^2 S(\omega) t.
\end{align}
%\subsection{The absence of run-to-run variance due to magnetic field gradients.}
%The capture signal rate in the nEDM@SNS undergoing CSD is approximated by 
%\begin{align}
%    S(t)=\frac{e^{-t/\tau_{n3}}}{\tau_{n3}}\left(1-\left<\sigma_n\cdot\sigma_3\right>\right).
%\end{align}
%Where $\tau_{n3}$ is the harmonic average of the neutron-$^3$He capture rate and the neutron lifetime $\tau_n$. In the presence of magnetic field gradients we can compute $\left<\sigma_n\cdot\sigma_3\right>$ with the wave functions from reference \cite{swank}. Now in the Redfield limit where the position auto-correlation decay time of the particles is much shorter than the coherence time ($\tau_c<<T_2.$) we have
%\begin{align}
%\left<\mathbf{\sigma_n}\cdot\mathbf{\sigma_3}\right>=\left<\mathbf{\sigma_n}\right>\cdot\left<\mathbf{\sigma_3}\right>,
%\end{align}
%and furthermore,
%\begin{align}
%    \left<\mathbf{\sigma_n}\right>\cdot\left<\mathbf{\sigma_3}\right>=\left<\sigma_{zn}\right>\left<\sigma_{z3}\right>+\mathrm{Re}\left(\left<\sigma_{+n}\right>\left<\sigma_{-3}\right>\right).
%\end{align}
%We write a generic perturbed state $\ket{s}$ as
%\begin{align}
%    \ket{s}=\binom{a}{b}. \label{eq:gen2state}
%\end{align}
%TODO: Probably this could be pared down even more.
Lastly, we find an absence of run-to-run signal variance for neutrons and $^3$He, $\Var_{cl}(\sigma_n \cdot \sigma_3)$, from magnetic fluctuations arising from Brownian motion in a static magnetic gradient of any field along any direction, as long as the field does not change from run to run. This is due to two effects: first, for each run $\sigma_n \cdot \sigma_3$ is averaged over a large number of particles where each neutron and $\ce{^3He}$ samples a different magnetic field; and second, the neutron and $\ce{^3He}$ trajectories are uncorrelated, and so equation \ref{eq:QMvariance} does not apply. Instead, because these trajectories are uncorrelated, the decay rate of the signal during a run can be computed from the relaxation rate of each individual species using the formalism of reference \cite{swank}. The relaxation rate and frequency shifts for each species are given by the expectation value of $\sigma_+$:
%An ensemble of neutrons and $^3$He undergoing stochastic motion will acquire a phase due to magnetic gradients in any field direction along any direction. Relaxation and frequency shifts due to arbitrary magnetic field gradients in spin dressed systems were described in reference \cite{swank}, and were derived by evaluating the expectation value of $\sigma_+$ from a wave function that contains terms up to second-order in magnetic gradients of any field direction along any direction, we find that
\begin{align}
\mathrm{Re}(\left<\sigma_+\right>)=1-\frac{t}{T_2}, \\
\mathrm{Im}(\left<\sigma_+\right>)=\delta\omega t.
\end{align}
%The trajectories of the neutron and $^3$He are uncorrelated. 
%Thus, if we apply the formalism from reference \cite{swank} to compute $\left<\sigma_n\cdot\sigma_3\right>$ none of the cross terms between $\ket{s_n}$ and $\ket{s_3}$  will contribute because the different species are uncorrelated.
Therefore, keeping only terms to second order in magnetic gradients after computing the expectation value of $\sigma_+$ and $\sigma_z$ with the perturbed wave function we find
\begin{align}
    \mathrm{Re}(\left<\sigma_{+_n}\right>\left<\sigma_{-_3}\right>)&=1-\frac{t}{T_{2_n}}-\frac{t}{T_{2_3}},\\
    \left<\sigma_{z_n}\right>\left<\sigma_{z_3}\right>&\approx0.
\end{align}
%because trajectories of neutrons and $^3$He are uncorrelated there is no contribution to the run-to-run signal variance from unchanging magnetic field gradients. 
Using the identity $\sigma_n \cdot \sigma_3 = \sigma_{z_n}\sigma_{z_3} + \Re(\sigma_{+_n}\sigma_{-_3})$, we derive
%Magnetic field gradients in any field direction along any direction contribute to the relaxation rate of the signal which can be approximated by,
\begin{align}
    \frac{d\left<\sigma_n\cdot\sigma_3\right>}{dt}\frac{1}{\cos(\phi_n-\phi_3)}=-\left(\frac{1}{T_{2_n}}+\frac{1}{T_{2_3}}\right) \approx -\frac{1}{T_{2_3}}.
\end{align}
Transverse relaxation, $T_2$, for both species can be calculated from reference \cite{swank}. Typically, because the trajectories of UCN are ballistic, and highly oscillatory in the measurement cell resulting in significant motional narrowing, the contribution of the transverse relaxation of the neutron to the signal decay is small. This result shows that there is no run-to-run phase fluctuation if gradients of all magnetic fields remain the same run-to-run. Gradients in magnetic fields diminish the statistical sensitivity by decreasing the coherence time of a single measurement, and the effect is predictable without stochastic variations contrary to the case with dressing field amplitude fluctuations.
\subsection{The feasibility of dressing with a cosine waveform.\label{sec:cosvsin}}
In order to minimize the variance in equation \ref{eq:QMvariance}, both spins should start in the plane of precession at $t=0$. However, in deriving equation \ref{eq:QMvariance} we assumed a magnetic field with a cosine dependence, i.e. $B_x(t) = B_1 \cos(\omega t)$. This poses a problem, as such a dressing pulse would require an instantaneous change in magnetic field at $t=0$, which is impossible to achieve due to the inductance of the dressing coil and limited slew rates of power supplies. Instead, the same dynamics can be achieved with sine dressing, i.e. $B_x(t) = B_1 \sin(\omega t)$, by having the neutron and $\ce{^3He}$ spins start outside the plane of precession. The polar angle $\theta_{n,3}$ and absolute phase $\phi_{n,3}$ of neutron and $\ce{^3He}$ required at $t=0$ to achieve this are specified in table \ref{table:startingphase}.

\begin{table}
\centering
\begin{tabular}{lllll}
\hline
\textbf{$\phi_{n3}$}  & \textbf{$\phi_n$}  & \textbf{$\theta_n$}  & \textbf{$\phi_3$}  & \textbf{$\theta_3$}  \\ \hline
0 & -0.99650 & 1.81020 & -1.00759 & 1.85511 \\ 
$\pi/4$ & -0.59186 & 1.84162 & -1.40374 & 1.77373 \\ 
$\pi/2$ & -0.18485 & 1.83092  & -1.78485 & 1.66284 \\ \hline
&  &  &  &
\end{tabular}%
\caption{Optimized parameters for the starting position on the Bloch sphere for neutrons and $^3$He to achieve cosine dressing with a sine dressing field described in section \ref{sec:cosvsin}. Where $\phi_{n3}=\phi_{n}-\phi_{3}$ is the relative azimuthal angle of the neutron and $^3$He that is desired during critical dressing, $\phi_{n,3}$ and $\theta_{n,3}$ are the absolute azimuthal angle and absolute polar angle of the neutron and $^3$He respectively. When the neutron and $^3$He start at these position with sine dressing the result is cosine dressing dynamics for the respective relative phase $\phi_{n3}$. All units are in radians. Values are for the fiducial nEDM@SNS experiment parameters for critical dressing, where the dressing field amplitude is 40.2497 $\mu$T oscillating at 1 kHz with a holding field strength of 5.2 $\mu$T, optimization of the fiducial nEDM@SNS parameters are discussed in reference \cite{nEDMSNS}.}
\label{table:startingphase}
\end{table}

%%%%%%%%%%%%%%%%%%%%%%%%%%%%%%%%%SIMULATIONS
\section{Simulation}
\begin{figure}
    \centering
    \includegraphics[width=\columnwidth]{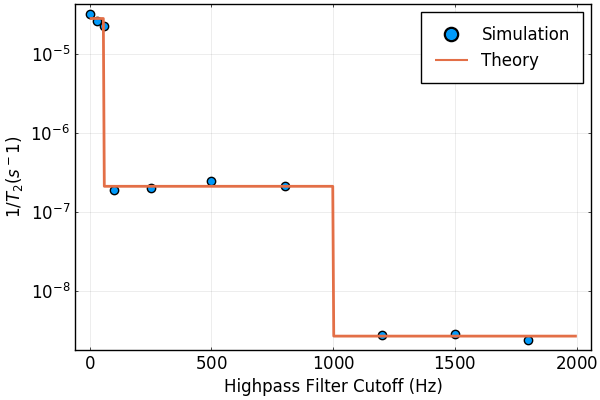}
    \caption{Predicted and simulated relaxation rates for various highpass filter cutoffs. Each simulated point represents 200 neutrons. Note that cutoff values beyond the dressing frequency (1000 Hz) are only achievable in simulation, and are shown here for comparison to the theory.}
    \label{fig:noise_vs_frequency}
\end{figure}

\begin{figure}
    \centering
    \includegraphics[width=\columnwidth]{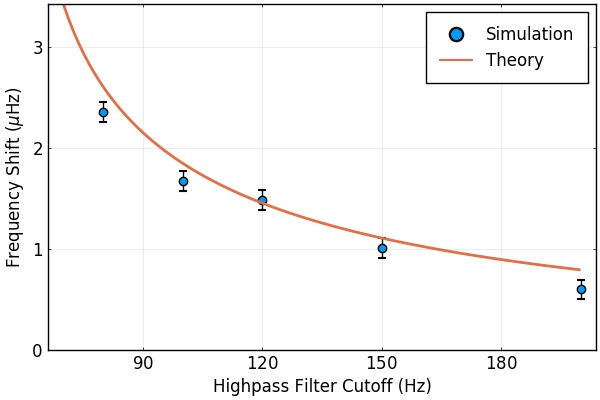}
    \caption{Predicted and simulated frequency shift for various highpass filter cutoffs. Each simulated point represents 100,000 neutrons.}
    \label{fig:shift_vs_frequency}
\end{figure}

The calculations in the previous section are verified by numerically integrating the Bloch equations with an adaptive Runge-Kutta integrator. Ensembles of spins are simulated using the fiducial spin dressing parameters for the nEDM@SNS experiment ($B_1\approx \SI{40}{\micro\tesla}$, $\omega\approx\SI{2\pi}{\kilo\hertz}$), with the addition of a noisy magnetic field with amplitude spectral density of \SI{5.7e-7}{G/\sqrt{Hz}}. The noise power spectrum is varied by changing the cutoff frequency of a simulated highpass filter. Figure \ref{fig:noise_vs_frequency} shows the simulated and theoretical relaxation times for different values of highpass filter cutoff. As evidenced by a sharp decrease in relaxation rate as the cutoff frequency is swept pass $2\pi\omega_0'$ and $2\pi\omega$, $T_2$ depends only on the power spectral density at a set of discrete frequencies. In the specific case of magnetic field fluctuations due to current fluctuations in the dressing coil, a physical highpass filter in series with the coil can reduce noise at $\omega_0'$. However, this is not possible for noise at $\omega$, which correspond to amplitude fluctuations of the dressing field. Thus, in the following sections we will explore several techniques to mitigate this type of noise. Figure \ref{fig:shift_vs_frequency} compares the simulated and theoretical frequency shifts. The frequency shift decreases as the highpass filter cutoff is increased past the dressed Larmor frequency. Therefore if a physical highpass filter is employed to reduce the run-to-run phase variation due to current fluctuations, its cutoff frequency should be chosen to be as high as possible. It should be noted however that for the case of nEDM@SNS experiment, a frequency shift caused by current fluctuations can be corrected for by comparing the signal from the two measurement cells for the same run.

\section{Mitigation Strategies}
\subsection{Feedback Control}
\begin{figure}
    \centering
    \includegraphics[width=\columnwidth]{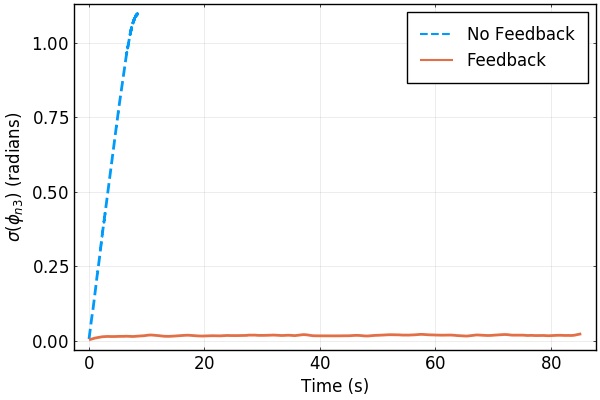}
    \caption{The root mean squared deviation of the angle between the neutron and \ce{^3He} spins as a function of time, for magnetic fields generated with (solid line) and without (dashed line) feedback control. The dataset without feedback consists of 28 runs, while the dataset with feedback consists of 50. Data are smoothed for clarity.}
    \label{fig:feedback_both}
\end{figure}

\begin{figure}
    \centering
    \includegraphics[width=\columnwidth]{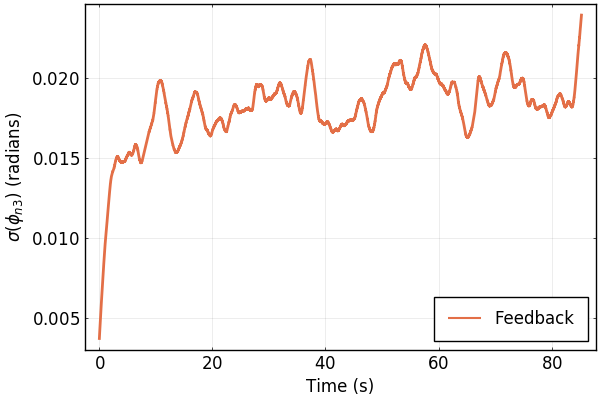}
    \caption{Phase standard deviation with feedback control applied. The rms phase deviation rises for a period of less than five seconds, then appears to settle near a fixed value. Data are smoothed for clarity.}
    \label{fig:feedback_only}
\end{figure}

The considerations in the previous sections show that fluctuations in the nEDM signal due to power supply instability would arise primarily due to the noise power spectrum at $S(\omega)$, which corresponds to amplitude fluctuations in the dressing field. In particular, the phase accumulated between the neutron and $^3$He is proportional to the integral of the dressing field amplitude over time. Thus, one way to limit the impact of power supply noise would be to employ feedback control to minimize the deviation of this integral from a target value. Using a Kepco four-quadrant 400 36-12 power supply and a DT9837A digital signal analyzer, we demonstrate a proof-of-concept for using feedback control in the nEDM@SNS experiment. The signal analyzer outputs a 1000 Hz sine wave, which we feed into the voltage control input of the power supply. The power supply drives a cosine coil, at the center of which is a pickup coil. The voltage across this pickup coil is measured by the signal analyzer, and the output of the signal analyzer is adjusted based on this measurement using a digital PI loop to maintain the integral of the signal amplitude near its target value. We then use the measured fluctuating magnetic field to simulate the motion of neutron and $^3$He spins, using the CODATA values for their gyromagnetic ratios \cite{NISTCODATA}. The measured magnetic field is postprocessed with a highpass filter to remove lower-frequency noise, and a single global scaling factor is applied to the field to replicate the CSD parameters of the nEDM@SNS experiment. Figures \ref{fig:feedback_both} and \ref{fig:feedback_only} show the resulting rms deviation of the relative phase shift between neutron and $^3$He spins.


\subsection{Cross-Cell Correlation}
In order to combat systematic effects, the nEDM@SNS experiment uses two measurement cells which will be exposed to opposite electric fields. Under the assumption that the magnetic field noise $\delta B(t)$ is identical in the two cells, it may be possible to correlate the effects of dressing field noise in the two cells and thus eliminate its effect on the measurement of the neutron EDM. One confounding factor for this strategy is the presence of static magnetic field gradients in the holding field. Due to imperfections in the the holding and dressing field coils, the dressed Larmor frequencies between the two cells may differ by a small amount. In this case, the same noise field $\delta B(t)$ may affect the two cells differently, hampering our ability to correlate the two cells. In order to determine the correlations between the spins in different cells, we compute the expectation value $\expval{\vec{\sigma}_1 \cdot \vec{\sigma}_2}$, where $\vec{\sigma}_1$ and $\vec{\sigma}_2$ represent two spins of the same species in opposite cells. Let $\omega_1$ and $\omega_2$ be the dressed Larmor frequencies in the two cells. In this case, calculating the expectation value is complicated by the fact that $\omega_1$ and $\omega_2$ may be very close in frequency, and thus we can no longer use the approximation that cross-frequency terms may be neglected. These terms are calculated in the appendix (section \ref{sec:near frequency}). The expectation value $\expval{\vec{\sigma}_1 \cdot \vec{\sigma}_2}$ is then given by
\begin{equation}
    \expval{\vec{\sigma}_1 \cdot \vec{\sigma}_2} = 1 - \frac{\gamma^2}{2} S(\bar{\omega}) \left(t - \frac{\sin(2 \Delta \omega t)}{2 \Delta \omega} \right)
\end{equation}
We see that decorrelation between the two cells may be neglected if either $S(\bar{\omega})$ is small, or $\Delta \omega t$ remains small over the course of the experiment.

\section{Robust Dressing}
We turn our attention to strategies to amplify coherence times of systems undergoing fluctuating fields. With specially tuned dressing modulation we find we can significantly decrease the decoherence, this technique is similar to those found in references \cite{KazakovMagicDressing2015,BoothMultiDressing2018, bevilacqua2021harmonic}. Here we apply modulation with angular frequency $\omega_m$ to the dressing parameter $x=\gamma B_1/\omega$, where $\omega_m$ is faster than the dressed Larmor precession $\omega_0'$. This parameter is tuned in conjunction with dressing parameters $B_1$ and $\omega$ so there is no phase accumulation. In a system with two spin species this can be tuned such that both states on a short time scale are rapidly oscillating but on average are effectively frozen in time. With this combination of fast oscillation and no phase accumulation, quantum states become especially robust to field fluctuations. On average both spin species remain fixed on the Bloch sphere for arbitrary long timescales limited by the coherence, this technique may prove useful in storing quantum information by keeping the state static and robust to electromagnetic fluctuations.  

Robust dressing is a valuable technique for the nEDM@SNS experiment, this experiment is performed in a large cryogenic apparatus. Modifications to the apparatus require a long dead time to allow the apparatus to warm, and then cool to operating temperature (T$<$500 mK). If there exist any unintended holding field gradients in the precession volume, gradients can be determined by a cryogenic probe array described in reference \cite{Nouri14}, or by  measuring transverse relaxation rates versus the applied field, and then shimmed away. However, during the initial phases of the experiment there may exist holding field gradients large enough that the scintillation signal decays too quickly to be measured effectively. Such gradients may be caused, for example, by an unknown nickel flash (nickel coating) on an electric connection close to the measurement cell. Furthermore, in this particular spoiled field scenario, the probe array described in reference \cite{Nouri14} will not be sufficient to reconstruct the field due to the reconstruction condition that there are no magnetic sources within the volume of interest. In such cases, a modulated spin dressing technique can be employed to increase the coherence time by several orders of magnitude. 

In the following sections we first outline a form of precession-free dressing which we call robust dressing because it is particularly robust against relaxation and dressing field noise. This is similar to results obtained for multiple field dressing in references \cite{Bevilacqua2020,bevilacqua2021harmonic}. While these references consider dressing fields acting along multiple axes, here we limited ourselves to a single dressing field in the $\hat{x}$ direction, as is the case for nEDM@SNS experiment. We then apply this technique to find precession-free dressing for both \ce{^3He} and neutron simultaneously, which we call robust critical dressing (RCD). 

\subsection{Description within the Bessel function approximation}
The dressed system of a single species can be held at a value where no phase accumulates, around the zero crossing of the Bessel function, $J_0(\gamma B_1 / \omega)\approx0$. Alternatively, it can be modulated between dressing parameter values that achieve a system where no phase accumulation occurs. For example, the effective absolute phase of the neutron is approximately
\begin{align}
\phi_n(t)&=\int_0^t\gamma_n J_0\left(\frac{\gamma_n B_1(t)}{\omega}\right)B_0dt.
\end{align}
For example, if the phase of a spin, in a plane perpendicular to the holding field, undergoing RCD modulation is given by
\begin{align}
\phi_n(t)&=\int_0^t\gamma_n J_0\left[x_0+\frac{x_1}{2}\left(1+\cos\omega_mt\right)\right]B_0dt. \label{eq:RCDcosmod}
\end{align}
then robust dressing is achieved when
\begin{align}
    \left< \phi_n(t)\right>=0, \label{eq:zeroPhase}
\end{align}
which can be found numerically. The parameters required to achieve robust dressing in simulations with cosine modulation can be estimated from equation (\ref{eq:RCDcosmod}) and (\ref{eq:zeroPhase}) by integration of the modulation function over a modulation period; moderate deviations from the values predicted analytically are expected due to violation of the Landau-Zener approximation. This is described in the context of spin dressing in reference \cite{EckelDressing2012}. It is found that the shorter the modulation period the further the RCD parameters deviate from the analytic prediction. In the simple analytic formulation it is found that if $x_0=1.2$ and $x_1\approx3.17$ then $\left< \phi_n(t) \right>\approx0$ (for both the neutron and $^3$He.)
Despite violation of the Landau-Zener approximation, detailed simulations of the Bloch equation find a continuum of values for $x_0$ and $x_1$ that satisfy the robust dressing condition,$\left< \phi_n(t)\right>=0.$
Deviations away from the average phase can be made small when $\omega'_0 < \omega_m$. The average projection of the spin on the Bloch sphere remains fixed, with fluctuations on the order of the modulation period and the instantaneous precession rate.

\subsubsection{Robustness}
Robust dressing suppresses transverse relaxation due to field fluctuations in $B_z$ and $B_y$. For $B_z$, this can be seen from the dressing approximation - the effective gyromagnetic ratio under precession-free dressing is zero, and therefore the local holding field strength is irrelevant. Likewise, a variation in $B_y$ can be viewed as a rotation of the holding field, which in the case where the effective gyromagnetic ratio is zero has no effect on the spins' overall behavior. Thus, the only static gradients and field offsets which contribute to the transverse relaxation rate are those in $B_x$ along any direction. To analyze these it is useful to consider a simplified model of robust dressing.

\subsection{An Intuitive Model of Robust Dressing}
While we are not aware of any purely analytical solutions for this type of pulse, robust dressing has several properties that allow for a convenient approximation. We have chosen the modulation frequency $\omega_m$ such that it evenly divides the dressing field frequency $\omega$. Thus, the pulse is periodic with frequency $\omega_m$, and its behavior is entirely defined by its action on a spin during the interval $t=0$ to $t=2\pi/\omega_m$. This property allows us to analyze this pulse using Floquet theory, described in reference \cite{BaroneFloquet}. In short, Floquet's theorem allows us to write the time evolution operator of robust dressing as
\begin{equation}
    T(t) = M(t) e^{-i \Lambda t},
\end{equation}
where $M(t)$ is a unitary operator with period $2\pi/\omega_m$ and $\Lambda$ is a time-independent Hermitian operator. The operator, $\Lambda$, can then be written in the form
\begin{equation}
    \Lambda = -\frac{1}{2} \gamma \vec{B_{\text{eff}}} \cdot \vec{\sigma}.
\end{equation}
The overall rotation under robust dressing can thus be treated as arising due to an effective magnetic field $\vec{B_{\text{eff}}}$. The effective magnetic field,  $\vec{B_{\text{eff}}}$, can be computed by numerical integration of the Bloch equations over a single period, or approximated perturbatively as in references \cite{Bevilacqua2020} and \cite{bevilacqua2021harmonic}.

%This is similar to the techniques and findings in \cite{Bevilacqua2020} and \cite{bevilacqua2021harmonic}. Specifically in reference \cite{bevilacqua2021harmonic} they use dual strong dressing fields to find solutions for zero effective fields in all directions, they also find that the effective Larmor precession can be increased. Here we are limited to the modulation of a single dressing field and are motivated to find applications in permanent electric dipole moment searches and gradient field metrology. However, we find similar phenomena, for example, robustness to static field gradients resulting in relaxation and frequency shifts.

Numerically, we find that robust dressing corresponds to a weak ($|\gamma B_{\text{eff}}| \approx \text{1 Hz}$) magnetic field whose primary component is in the $\hat{x}$ direction. The precise orientation of $\vec{B_{\text{eff}}}$ may be adjusted by careful selection of the robust dressing parameters, and we find that this modulation technique is a useful form of quantum control. 

\begin{figure}
    \centering
    \includegraphics[width=\columnwidth]{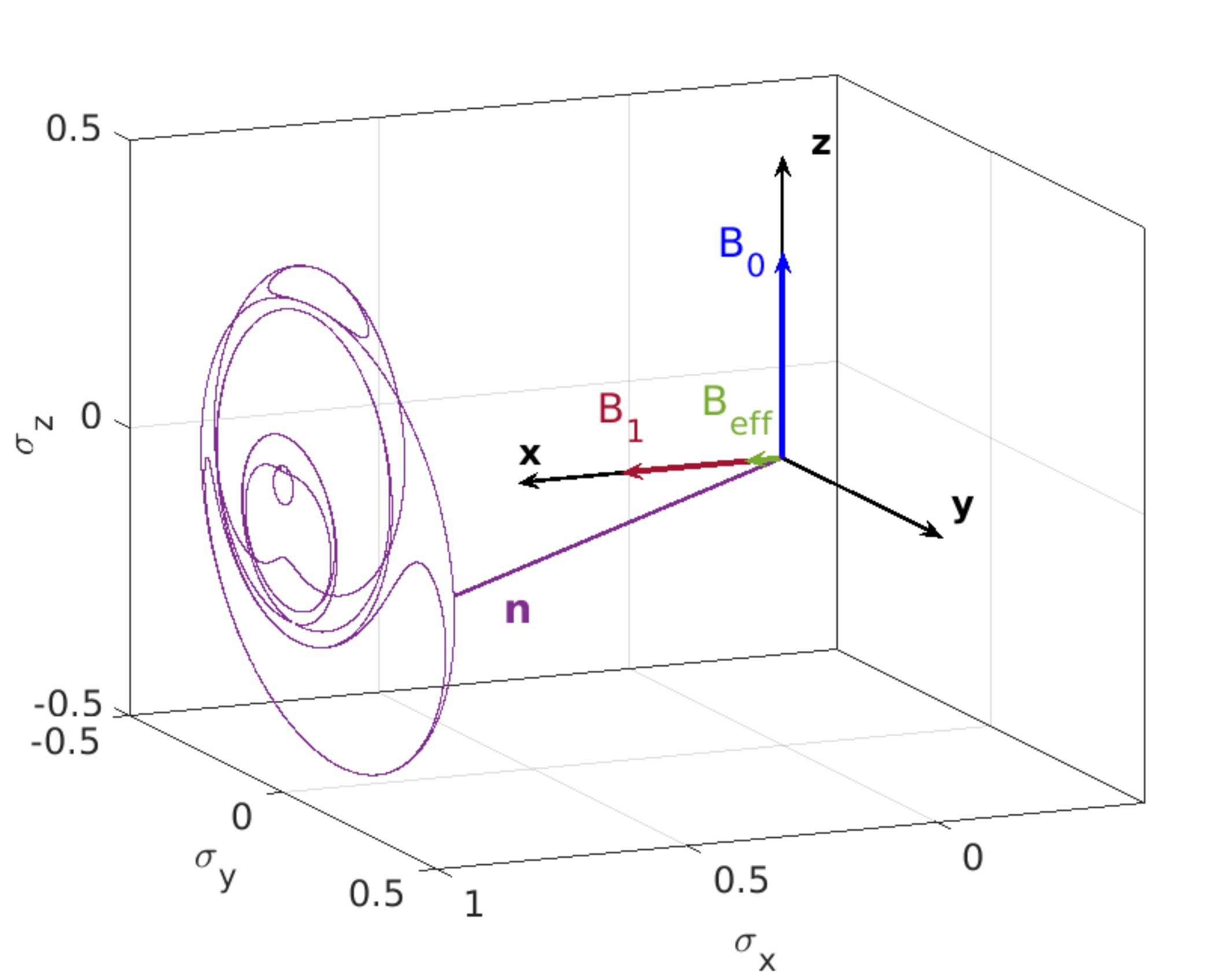}
    \caption{A classical visualization of the neutron spin trajectory undergoing robust dressing on the Bloch sphere, with arrows representing the holding field, $B_0$, the dressing field, $B_1$, and the robust dressing effective field $B_{\text{eff}}$.}
    \label{fig:RCDSchem}
\end{figure}

In light of this, we analyze relaxation in the robust dressing scenario by treating $\vec{B_{\text{eff}}}$ as a holding field which lies along $\hat{x}$ and redefining $T_1$ and $T_2$ accordingly. We also define $\phi$ to be the angle between the spin and $\hat{x}$. A geometrical representation of the robust dressed system for a neutron is shown in figure \ref{fig:RCDSchem}.

\subsection{Robust dressing applications}
Before presenting an analysis of relaxation under robust dressing, we describe several applications of robust dressing for an nEDM search such as nEDM@SNS experiment.
\subsubsection{Gradient metrology through spin relaxation}
Because the effects of gradients in $B_y$ and $B_z$ along any direction are suppressed by robust dressing, $T_2$ is always much shorter than $T_1$ for the nEDM@SNS experiment's expected operating and initial commissioning conditions. This will lead to a fast initial decay dominated by $T_2$, followed by a slower decay whose rate is given by $T_1$. In general, the polarization of spins decays as 
\begin{align}
P(t)=\sqrt{e^{-\frac{2t}{T_1}} \cos^2\phi_0 + P^2_{T_2}(t) \sin^2 \phi_0},
\end{align}
where  $P_{T_2}(t)$ is the component of the polarization transverse to $\hat{x}$ and $\phi_0$ is the value of $\phi$ at the onset of robust dressing. The transition between fast and slow decay then occurs at polarization $P_0 \approx \cos(\phi_0)$. This behavior can be applied for gradient metrology, as a gradient in $B_x$ will cause the spins to relax quickly to a known value, while robust dressing extends the coherence of the system from gradients in $B_y$ or $B_z$.  For the nEDM@SNS experiment, a static magnetic gradient of any field transverse to the holding field direction along the dressing axis represents the largest potential systematic effect by generating a frequency shift linear in the applied electric field, as evidenced by references \cite{commins,pendlebury04,Lam05,barabanov,clayton,Swank12,pignol,swank2016,Swank2018}. If this precise gradient measurement technique is used to feedback on magnetic gradient shimming, a powerful mitigation strategy for this linear-in-E frequency shift is achieved.

\subsection{Relaxation under robust dressing}
The longitudinal relaxation of the system can be decomposed into two contributions. One contribution arises from the Redfield-like longitudinal decay, which is generated from field imperfections in $B_y$ and $B_z$, and which we compute by evaluating the system in the interaction picture, similar to the dual harmonic dressing treatment provided by reference \cite{bevilacqua2021harmonic}, but with the oscillating field in only the $\hat{x}$ direction. A derivation is found in appendix \ref{sec:robustRelaxation}. The model predicts relaxation to be extended beyond the $T_1$ time found in reference \cite{redfield,mcgregor} for the corresponding holding field of the nEDM@SNS experiment's operating parameters, discussed in reference \cite{nEDMSNS}. This is because the integral over the field correlation function is shifted in frequency according to the harmonics of the dressing field and not the holding field as is the case in reference \cite{redfield}. An expansion of the spin's reaction to the applied field into a Fourier series is used to formulate the longitudinal relaxation rate. In the diffusion limit we find 
\begin{align}
\frac{1}{T_1}\approx \gamma^2 \left(G_y^2+G_z^2\right) D \sum_n \frac{|a_n|^2}{n^2 \omega_m^2}, \label{eq:rcdrelax}
\end{align}
where $a_n$ are the Fourier coefficients of the series and are found numerically. This result can be extended beyond the diffusion limit by using the spectrum of the trajectory autocorrelation function; this is described in appendix \ref{sec:robustRelaxation}. 

An additional source of longitudinal relaxation arises from the field imperfections in $B_x$, which generate longitudinal relaxation by rotating the local effective field off of the $x$ direction, resulting in a contribution of the effective field along $y$ and/or $z$. This in turn, will cause relaxation according to
\begin{align}
    \frac{1}{T_1} &\approx \left(G_{\perp, \text{eff}}\right)^2  \frac{D}{\abs{B_{0, \text{eff}}}^2} \label{eq:T1eff} , \\
    &\approx \gamma^2 \left(\pdv{B_{\perp,\text{eff}}}{B_x}\right)^2 G_x^2 \frac{D}{\abs{B_{0, \text{eff}}}^2},
\end{align}
where $G_{\perp,\text{eff}}$ and $B_{\perp,\text{eff}}$ refer to components of the effective gradient and magnetic field orthogonal to the effective holding field $B_{0, \text{eff}}$. Validity of this estimate depends on the magnitude of gradients in $B_x$ in the cell. In particular, if $L_x$ is the length of the cell along the x direction, then this approximation breaks down when $\abs{G_x L_x/2}\gtrsim\abs{B_{\mathrm{eff}}}$ because the scaling of $B_{\perp,\text{eff}}$ with $B_x$ becomes non-linear. The longitudinal relaxation will not be proportional to $G_x^2,$ but will scale with a higher power of $G_x$ due to the non-linearity in the effective field scaling. To find the exact scaling in closed form is difficult. However, it is found from simulations that the robust dressing parameters can be tuned to mitigate this source of relaxation, even for large gradients in $B_x$. By tuning the robust critical dressing parameters to minimize the relaxation rate, we can mitigate the contribution from the $B_x$ field imperfections to, at least, the level of contribution from the other directions. In this optimized tuning the relaxation rate from this gradient returns to $G_x^2$ scaling, implying a return to the linear scaling of the effective field under optimized robust dressing. Due to the ability to mitigate or increase relaxation by tuning the robust dressing parameters we find that this modulation technique is a useful form of quantum control.

\begin{figure}
    \centering
    \includegraphics[width=\columnwidth]{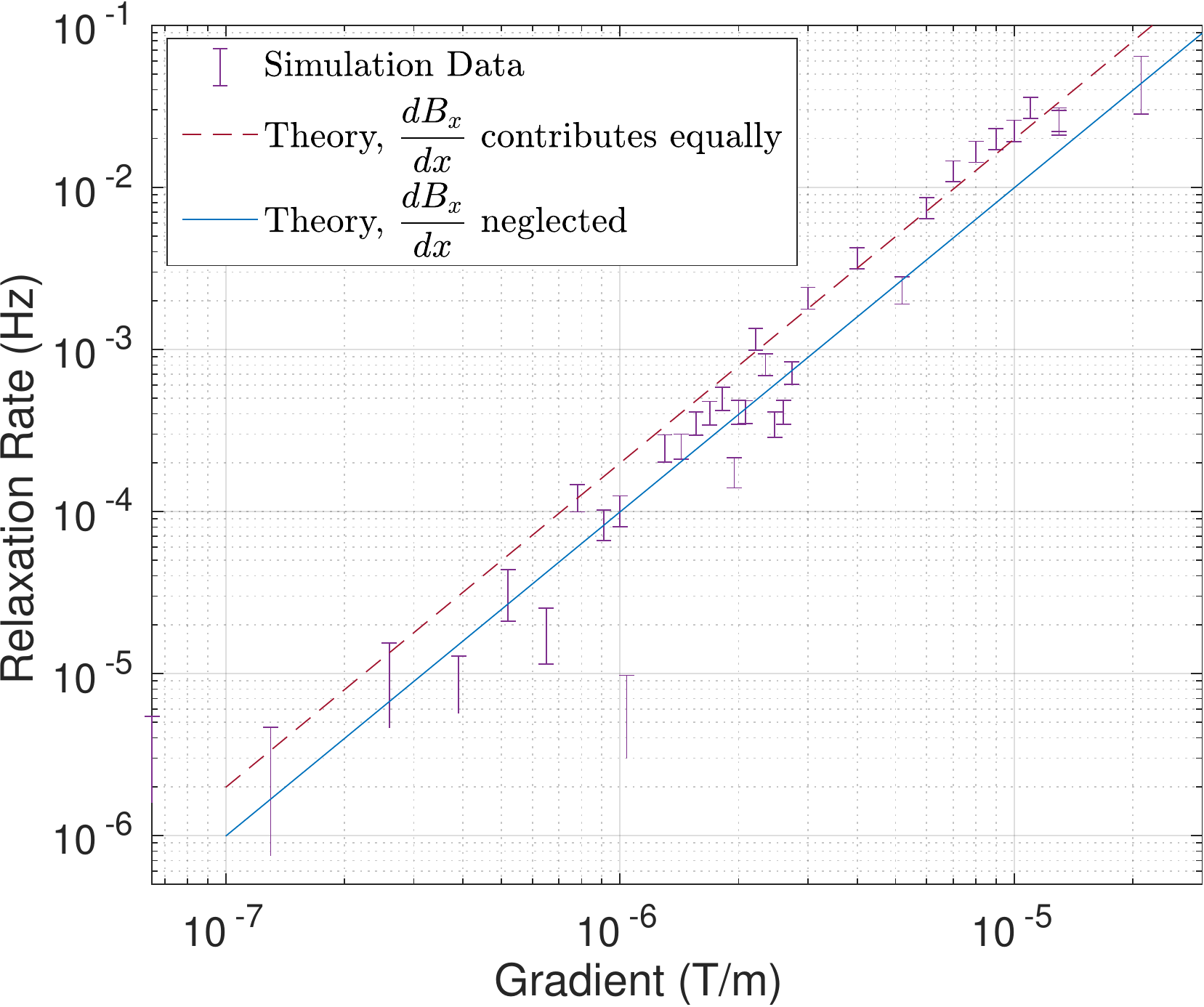}
    \caption{Relaxation after optimization of robust dressing parameters for 200 Hz robust dressing and a wide range of gradient magnitudes. The gradients in this simulation arise from the divergence theorem, $\frac{dB_z}{dz}=-\frac{dB_x}{dx}.$}
    \label{fig:robustDecay}
\end{figure}

In practice, because Maxwell's equations require a gradient contribution in at least two directions, this optimization procedure to minimize relaxation obtains diminishing returns when the relaxation from the gradient in $B_x$ becomes comparable in magnitude to the relaxation contributed from gradients in $B_y$ and/or $B_z$. Due to the nonlinear scaling, and the precision of the robust dressing parameters required, we expect that in the large gradient regime the relaxation from gradients in $B_x$ can be mitigated to parity with the contribution from $B_y$ and $B_z$. However, in the small gradient regime, where tuning need not be as precise, we may more easily find optimized robust dressing parameters where the contribution from the gradient in $B_x$ can be ignored. The prediction is formulated as 
\begin{align}
\frac{1}{T_1}=C_{dB_x} G_x^2+\gamma^2(G_y^2+G_z^2) D \sum_n  \frac{|a_n|^2}{n^2 \omega_m^2},
\end{align}
where $C_{dB_x}$ is determined from specifics of the parameters, but in general, it can be estimated to be of magnitude
\begin{align}
C_{dB_x} \leq \gamma^2 D \sum_n  \frac{|a_n|^2}{n^2 \omega_m^2}.
\end{align}
The relaxation from simulations of a linear gradient, where $G_x=dB_x/dx=-dBz/dz$, is shown in figure \ref{fig:robustDecay} for all of the regimes of robust dressing, as well as the theoretical prediction after optimization for both limits of $C_{dB_x}.$ The smallness of $B_\text{eff}$ compared to $B_0$ and the nonlinear dependence on $G_x$ lead naturally to three different regimes depending on the magnitude of the $G_x$ gradient relative to $B_\text{eff}$.

\subsubsection{Small gradient regime}
If $\abs{G_x L_x/2} \ll \abs{B_\text{eff}}$, the magnetic field gradient across the cell is smaller in magnitude than the effective holding field. The tuning of the parameters need not be terribly precise ($\delta x_{0,1}  \sim x_{0,1} \times 10^{-3}$) to achieve the $T_1$ relaxation time shown in equation \ref{eq:rcdrelax}. Additionally, the $G_x$ contribution to $T_1$ may be found using the effective magnetic field formalism using equation \ref{eq:T1eff}.

Similar to traditional NMR transverse relaxation the RCD analog can be predicted from references \cite{redfield,mcgregor}; however, only gradients in $B_x$ contribute significantly. Thus, for the transverse polarization we find
\begin{align}
P_{T_2}(t)=e^{-\frac{t}{T_2}},
\end{align}
where in the diffusion limit,
\begin{align}
\frac{1}{T_2}=\gamma^2 G_x^2\frac{ L_x^4}{120 D}.
\end{align}
Outside of the diffusion limit, the transverse relaxation can be found from the spectrum of the trajectory autocorrelation function at zero frequency. 
\subsubsection{Intermediate gradient regime}
If $\abs{G_x L_x/2} \sim \abs{B_\text{eff}}$, then the gradient magnetic field nearly cancels $B_{\text{eff}}$ in a significant portion of the cell. For untuned robust dressing parameters, the effective magnetic field may have a nonzero $B_{z,\text{eff}}$ component in this region, leading the local magnetic field here to be misaligned with the $x$-axis. This drastically increases the longitudinal relaxation rate for this gradient regime. While no theoretical prediction exists for the dynamics of the spins in this regime, a thorough investigation is presented in reference \cite{StollerHapperDyson}. The strong dependence of $T_1$ on $G_x$ in the vicinity of the intermediate gradient regime can be employed for gradient metrology. Specifically, a gradient large enough to induce this regime could be identified by measuring the $T_1$ relaxation of the spin species under study. Although the gradients in $B_x$ required to achieve this regime are a couple orders of magnitude larger than the initial gradients expected at the onset of commissioning of the nEDM@SNS experiment, they are in the realm of possibility if there is an unaccounted for magnetized material - for example, a nickel flash (nickel coating) on an electrical connection, or other unknown magnetic material close to the measurement cell.
Note that while precise ($\delta x_{0,1}  \sim x_{0,1} \times 10^{-5}$) tuning of the robust dressing parameters can eliminate the $B_{z,\text{eff}}$ component in the misaligned region, thus recovering the $T_1$ of equation \ref{eq:rcdrelax}, this is not required nor useful for gradient metrology.

\subsubsection{Large gradient regime}
As $\abs{G_x L_x/2}$ becomes large compared to $\abs{B_\text{eff}}$, the region of low magnetic field shrinks in volume, and the effect of misalignment becomes less significant. The treatment of $T_1$ and $T_2$ is straightforward. Longitudinal relaxation can be computed from equation \ref{eq:rcdrelax}. Transverse relaxation under gradients of this magnitude behave according to the adiabatic regime described in reference \cite{pignol2015}, as the spins dephase on a shorter timescale than the diffusion time. This results in a relatively fast relaxation, given by
\begin{align}
P_{T_2}(t)&=\frac{2}{\gamma G_x L_x t} \sin(\gamma G_x L_x t/2),\\
&=\mathrm{sinc}\left(\frac{\gamma G_x L_x t}{2}\right),
\end{align}
where $\mathrm{sinc}(x)=\frac{\sin(x)}{x}$.

\subsubsection{Robust dressing and dressing field noise\label{sec:RCDnoise}}
\begin{figure}
    \centering
    \includegraphics[width=\columnwidth]{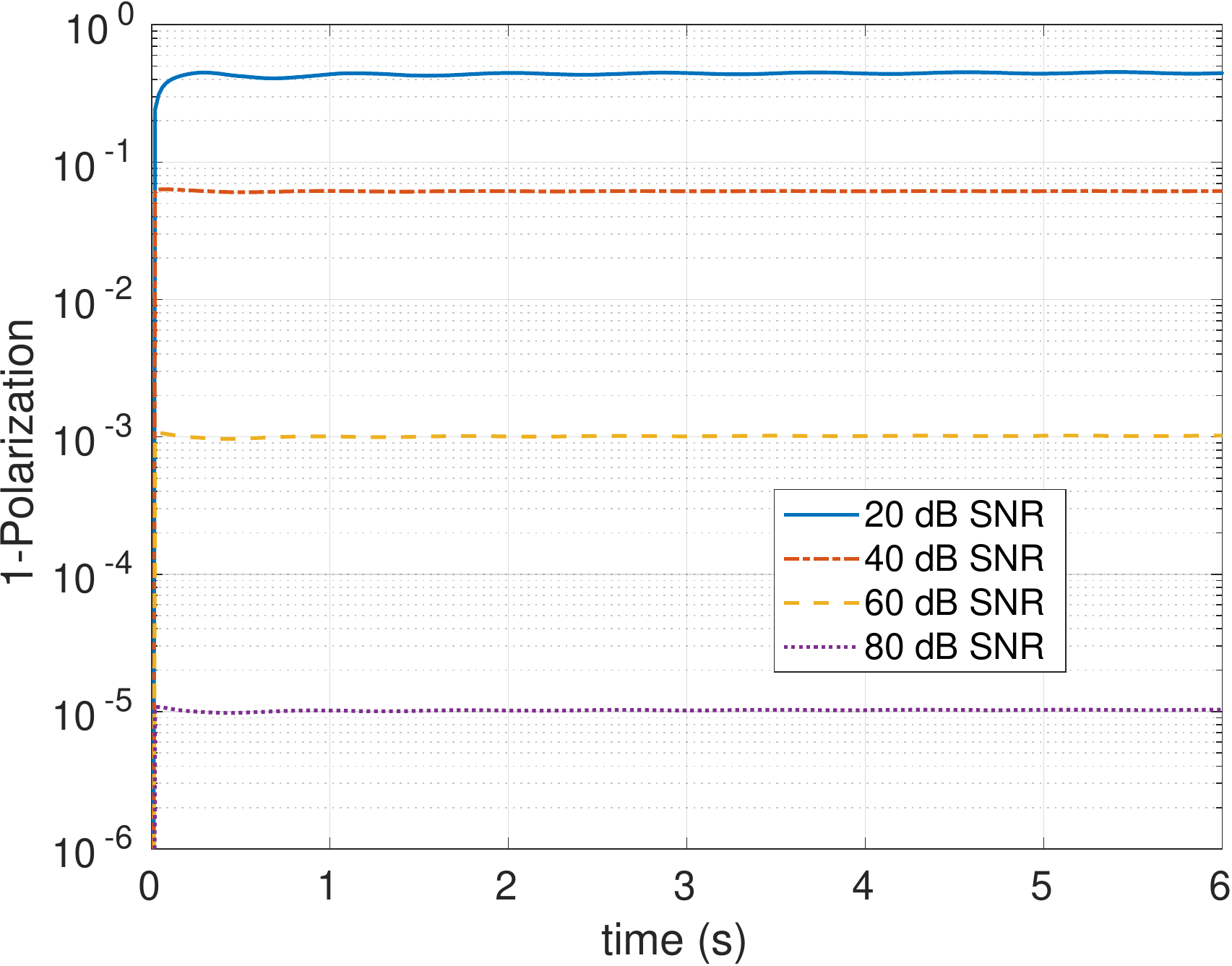}
    \caption{ 1- Polarization for robust dressing modulated at 200 Hz (red) with simulated noise as a parameter from 20 dB SNR to 80 dB SNR}
    \label{fig:robustNoise}
\end{figure}

From the simple model that describes robust dressing we expect that robust dressing is also robust to fluctuations of the dressing field, because the integral of the total phase accumulated on average is zero. In simulations of noise we find that after an initial relaxation period on the time scale of the modulation frequency, the random phase accumulation nearly ceases, and dressing field fluctuations arising from power supply noise can typically be ignored with commercial linear amplifiers. The polarization of a 200 Hz robust dressing scheme with amplitude noise from 20 to 80 dB SNR is shown in figure \ref{fig:robustNoise}. In that figure it is shown that at 20 dB SNR, where the amplitude of the field noise is 10\% of the amplitude of the pulse, polarization can be observed for a long time, and the rate in equation \ref{eq:rcdrelax} is still achievable after an initial loss. 

\subsubsection{Robust critical dressing}
There exists a non-trivial solution that satisfies the robust dressing conditions for both the neutron and \ce{^3He} at the same time. Robust critical dressing is achieved when
\begin{align}
\left<\phi_n\right>  \approx \left<\phi_3\right>\approx0. 
\end{align}
Table \ref{table:robust} shows optimized robust dressing parameters for the CODATA values of the magnetic moments of the neutron and $^3$He. We find that robust critical dressing significantly reduces relaxation from field fluctuations in the holding field and dressing fields.
%% This bit I'm not sure about
%%we have a whole section on this probably don't need this here, it is confusing.  How about  
%Furthermore, it is robust against fluctuations in the dressing field amplitude because it is quickly modulated between positive and negative effective couplings. 

\begin{figure}
    \centering
    \includegraphics[width=\columnwidth]{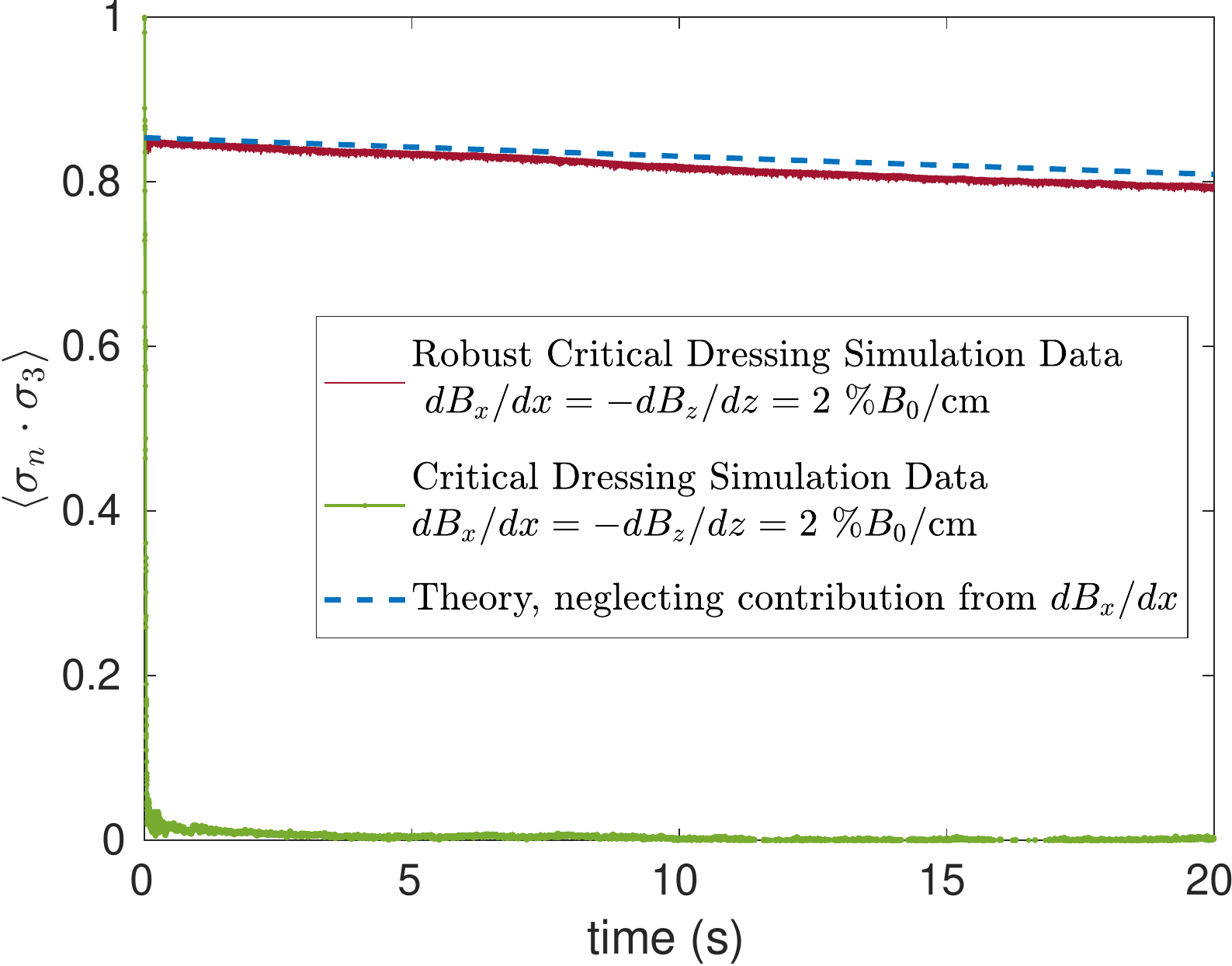}
    \caption{Simulated signal for highly tuned RCD modulated at 200 Hz (red) and CSD (green),  with theoretical estimation for RCD (blue).  Starting phase for $^3$He is $\phi_3=\pi/8$, and it is the same for neutrons, $\phi_n=\pi/8$. Signal is observable for a long time in the RCD scenario, while the signal is indistinguishable from noise in a traditional critical dressing scenario.}
    \label{fig:robust}
\end{figure}

Figure \ref{fig:robust} shows a simulation of the robust critical dressing signal for $\omega_m = 2\pi \times \SI{200}{\hertz}$ in a large gradient, along with the theoretical estimation for robust critical dressing based on
\begin{align}
\left<\sigma_n\cdot\sigma_3\right>\approx P_{0}^2 \exp\left( \gamma_3^2 G_z^2 D  t  \sum_n \frac{|a_n|^2}{n \omega_m}  \right).
\end{align}
We have ignored neutron relaxation in this prediction, as it is expected to be small because neutrons exhibit strong motional narrowing. The applied gradient satisfies Maxwell's equations with $dB_z/dz=-dB_x/dx=2 \times 10^{-2} \mathrm{B_0/cm}$. For comparison, unmodulated critical dressing is also simulated in the same gradient, and the signal is found to decay rapidly compared to the robust case.

\subsubsection{Mitigation of linear-in-E effects}
Nominally, for nEDM@SNS experiment, the electric field is applied parallel to the holding field. In this case robust dressing cannot be used to detect an nEDM. However, if the electric field is instead applied along $\hat{x}$, robust critical dressing can be employed to measure an nEDM and simultaneously mitigate linear-in-E frequency shifts arising from the $v \times E$ effect via the same mechanism that mitigates the longitudinal relaxation; the response of the spin precession system to magnetic field fluctuations is determined by the power spectrum of the noise evaluated at integer multiples of the modulation frequency rather than at the Larmor frequency. Due to the $1/\omega^2$ dependence of the trajectory correlation functions in this regime the contribution to the linear-in-E frequency shift is significantly mitigated.

\begin{table}
\centering
\begin{tabular}{llll}
\hline
\textbf{$\omega_m/2\pi$} (Hz) & \textbf{$x_0$} & \textbf{$x_1$} & $%
\sqrt{\langle\Delta\theta_{n3}^2\rangle}$ (rad) \\ \hline
200 & 1.226 & 3.497 & 0.68 \\ 
300 & 1.2920 & 3.3297 & 0.41 \\ 
500 & 1.5602 & 2.5910  & 0.25 \\ \hline
&  &  & 
\end{tabular}%
\caption{Optimized parameters for robust critical dressing. All values are calculated for a holding field of 5.2 $\mu$T and a nominal critical dressing field of 40.2497 $\mu$T oscillating at 1000 Hz.}
\label{table:robust}
\end{table}

\section{Conclusion}
In this work we developed a model to evaluate the effect of magnetic field noise on a critically dressed system of neutrons and $^3$He atoms in the case where the Larmor frequency is a non-negligible fraction of the dressing field frequency. Applying this model to the case of magnetic field fluctuations parallel to the dressing field, we find that the relaxation time is given by an infinite sum over the noise power spectrum evaluated at discrete frequencies, with higher frequency terms suppressed by higher-order Bessel functions. For the purpose of calculating the impact of current fluctuations in the dressing field coil, these results demonstrate that the primary concern is amplitude fluctuations in the dressing field. These results were verified by numerical integration of the Bloch equations. \\
With these results as motivation, we propose several strategies to increase coherence and sensitivity for systems of dressed spins. Feedback control on the dressing amplitude, for example, can substantially reduce the phase uncertainty in the presence of current fluctuations in the dressing field coil, while correlating the signal rate from opposing cells allows for rejection of noise sources which are common to both cells. We emphasize that while these strategies are proposed in the context of the nEDM@SNS experiment, these techniques are generally applicable to systems where precise control of dressed spins is required. Of particular interest is robust dressing, which sets the effective Larmor frequency of each species to zero by modulating the dressing field. This substantially reduces relaxation due to magnetic field fluctuations, and effectively freezes a state in time which is useful for extending the coherence of quantum information. Measuring the relaxation of a robustly dressed system allows for precise determination and feedback on gradients parallel to the dressing field, and thus can be applied as gradient metrology for fundamental physics measurements which require a uniform magnetic field. For the nEDM@SNS experiment, this effect can be leveraged to measure or shim relatively large spatial gradients in a critically dressed system of neutrons and $^3$He. Furthermore, robust dressing allows for mitigation of linear-in-E frequency shifts in EDM experiments where the electric field is chosen to be parallel to the dressing field. \\

\onecolumngrid
\appendix
\section{Near-Frequency Terms} \label{sec:near frequency}
Suppose $V_I(t)$ contains a pair of terms with frequencies $\omega_j$ and $\omega_k$ which are very close in frequency, such that $\abs{\omega_j - \omega_k} \sim 1/t$. Then the cross-frequency term can no longer be neglected. Define the integral
\begin{equation}
    v_{jk} \equiv \int_0^t dt' \int_0^{t'} dt'' e^{-i \omega_j t'} e^{i \omega_k t''} \expval{\delta B(t') \delta B(t'') }.
\end{equation}
Using equation \ref{eq:factor frequency}, $v_{jk}$ can be written in terms of the noise power spectrum as 
\begin{align*}
    v_{jk} &= \frac{e^{-2i \Delta \omega t} - 1}{-2 i\Delta \omega} \left(\frac{1}{2} S(\bar{\omega}) -\frac{i}{2 \pi} \int_{-\infty}^{\infty} d\omega' \frac{S(\omega')}{\omega' - \bar{\omega}} \right).
\end{align*}
Then the cross terms of $T^\dagger A T$ are

\begin{align*}
    &\int_0^t \int_0^t dt' dt'' (Q_j e^{-i\omega_j t''} + Q_j^\dagger e^{i \omega_j t''}) A (Q_k e^{-i\omega_k t'} + Q_k^\dagger e^{i \omega_k t'}) \\
    &- \int_0^t dt' \int_0^{t'} dt'' (Q_j e^{-i\omega_j t''} + Q_j^\dagger e^{i \omega_j t''}) (Q_k e^{-i\omega_k t'} + Q_k^\dagger e^{i \omega_k t'}) A \\
    &- 
    \int_0^t dt' \int_0^{t'} dt'' A (Q_k e^{-i\omega_k t'} + Q_k^\dagger e^{i \omega_k t'})  (Q_j e^{-i\omega_j t''} + Q_j^\dagger e^{i \omega_j t''}) \\
    &+ (j \leftrightarrow k) \\
    &= Q_j A Q_k^\dagger (v_{jk} + v_{kj}^*) + Q_j^\dagger A Q_k (v_{kj} + v_{jk}^*) \\
    &- Q_j Q_k^\dagger A v_{kj}^* - Q_j^\dagger Q_k A v_{kj} \\
    &- A Q_k Q_j^\dagger v_{kj} - A Q_k^\dagger Q_j v_{kj}^* \\
    &+ (j \leftrightarrow k), \\
    &= v_{jk} (Q_j A Q_k^\dagger + Q_k^\dagger A Q_j - Q_k^\dagger Q_j A - A Q_j Q_k^\dagger) \\
    &+ v_{jk}^* (Q_k A Q_j^\dagger +  Q_j^\dagger A Q_k - Q_k Q_j^\dagger A - A Q_j^\dagger Q_k) \\
    &+ (j \leftrightarrow k), \\
    &= v_{jk} [Q_k^\dagger, [A, Q_j]] \\
    &+ v_{jk}^* [Q_k, [A, Q_j^\dagger]] \\
    &+ (j \leftrightarrow k).
\end{align*}

\subsection{Explicit Calculation for Case of Static Gradients}
We first consider noise near the dressed Larmor frequency $\omega_0'$. Using $Q_j = \frac{\gamma}{2} \sigma_{+, 1}$ and $Q_k = \frac{\gamma}{2} \sigma_{+, 2}$, we get
\begin{align*}
    \expval{\vec{\sigma}_1 \cdot \vec{\sigma}_2} &= 1 - \frac{\gamma^2}{4}(v_1 + v_1^* + v_2 + v_2^*) - \frac{\gamma^2}{4}(v_{kj}^* + v_{kj} + v_{jk} + v_{jk}^*), \\
    &= 1 - \frac{\gamma^2}{2} t S(\bar{\omega}) - \frac{\gamma^2}{4} \frac{\sin(2 \Delta \omega t)}{\Delta \omega} S(\bar{\omega}), \\
    &= 1 - \frac{\gamma^2}{2} S(\bar{\omega}) \left(t - \frac{\sin(2 \Delta \omega t)}{2 \Delta \omega} \right).
\end{align*}
Next, we consider the contribution of noise near $\omega$, the dressing field frequency. In this case, static gradients do not affect the time-dependence of the noise Hamiltonian - that is, the frequency of the complex exponential $e^{i\omega t}$ is unaffected by a change in $\omega_0$ or $B_1$ (see equation \ref{eq:W expanded}). Therefore there are no cross-frequency terms to consider, and so noise near $\omega$ which applies equally to both cells cannot cause the spins between the two cells to decorrelate.

\section{Transformation of the Signal Rate into the Interaction Picture}
Computing the variance of $\sigma_1 \cdot \sigma_2$ in the lab frame corresponds to computing the variance of $U_1 U_2 e^{iH t}  \sigma_1 \cdot \sigma_2 e^{-iHt} U_2^\dagger U_1^\dagger$ in the interaction picture. We begin by calculating $U_1 U_2 \sigma_1 \cdot \sigma_2 U_2^\dagger U_1^\dagger$.

From the definition of $U_1$,
\begin{align}
    U_1 &= D(\eta_1/2) \ket{+_x}\bra{+_x} + D^\dagger(\eta_1/2) \ket{-_x}\bra{-_x}, \\
    &= \frac{1}{2}D(\eta_1/2)(1 + \sigma_x) + \frac{1}{2}D^\dagger(\eta_1/2)(1 - \sigma_x).
\end{align}
The Pauli spin operators then transform as
\begin{align}
    U_1 \sigma_x U_1^\dagger &= \sigma_x , \\
    U_1 \sigma_y U_1^\dagger &= \frac{1}{2} (D(\eta_1) + D^\dagger(\eta_1))\sigma_y + \frac{1}{2} i (D(\eta_1) - D^\dagger(\eta_1)) \sigma_z , \\
    U_1 \sigma_z U_1^\dagger &= \frac{1}{2} (D(\eta_1) + D^\dagger(\eta_1))\sigma_z - \frac{1}{2} i (D(\eta_1) - D^\dagger(\eta_1)) \sigma_y .
\end{align}
So the observable $\sigma_1 \cdot \sigma_2$ transforms as
\begin{align}
    U_1 U_2 (\sigma_1 \cdot \sigma_2) U_2^\dagger U_1^\dagger &= (U_1 \sigma_{x1} U_1^\dagger) (U_2 \sigma_{x2} U_2^\dagger) + (U_1 \sigma_{y1} U_1^\dagger) (U_2 \sigma_{y2} U_2^\dagger) + (U_1 \sigma_{z1} U_1^\dagger) (U_2 \sigma_{z2} U_2^\dagger) \\
    &= \sigma_{x1}\sigma_{x2} \\ 
    &+ \frac{1}{2}(D(\Delta \eta) + D^\dagger(\Delta \eta)) (\sigma_{y1}\sigma_{y2} + \sigma_{z1}\sigma_{z2}) \\
    &- \frac{i}{2}(D(\Delta \eta) - D^\dagger(\Delta \eta)) (\sigma_{y1}\sigma_{z2} - \sigma_{z1}\sigma_{y2}).
\end{align}
where $\Delta \eta = \eta_1 - \eta_2$.

Applying the approximation in equation \ref{eq:bessel-approx}, we have that
\begin{equation}
    D(\Delta \eta) \approx \sum_{n, q} J_q(x_1 - x_2) \ket{n+q}\bra{n}.
\end{equation}
For the fiducial nEDM@SNS experimental parameters, the dressing parameters of the neutrons and \ce{^3He} atoms are, respectively,
\begin{align*}
    x_n &\approx -1.184, \\
    x_3 &\approx -1.317.
\end{align*}
Because $|x_n - x_3|$ is significantly smaller than 1, the magnitude of $J_q(x_n - x_3)$ is small for $|q| > 0$. For example,
\begin{align*}
    J_0(x_n - x_3) &= 0.9956, \\
    J_1(x_n - x_3) &= 0.0662, \\
    J_2(x_n - x_3) &= 0.0022.
\end{align*}
We therefore make the approximation that
\begin{equation}
    U_1 U_2 (\sigma_1 \cdot \sigma_2) U_2^\dagger U_1^\dagger \approx \sigma_1 \cdot \sigma_2 .
\end{equation}
Next, we need to compute $e^{iH_U t} \sigma_1 \cdot \sigma_2 e^{-iH_U t}$, where we define $H_U = U_2 U_1 H U_1^\dagger U_2^\dagger$. We make use of the identity
\begin{equation}
    e^{iH_U t} \sigma_1 \cdot \sigma_2 e^{-iH_U t} = \sum_{n, n', s_1, s_1', s_2, s_2'} \widetilde{\ket{n, s_1, s_2}} \widetilde{\bra{n', s_1', s_2'}} \widetilde{\bra{n_1, s_1, s_2}} \sigma_1 \cdot \sigma_2 \widetilde{\ket{n', s_1', s_2'}} e^{i \Delta \omega t}, \label{eq:time dependent identity}
\end{equation}
where as before $\Delta \omega$ is the frequency difference between the states $\widetilde{\ket{n, s_1, s_2}}$ and $\widetilde{\ket{n', s_1', s_2'}}$.
We make two simplifications to this expression. First, we are concerned with the time-averaged signal, and therefore we ignore any terms where $\Delta \omega \neq 0$. Second, we consider only the leading-order terms in $\omega_0'/\omega$, and therefore make the approximation that $\widetilde{\ket{n, s_1, s_2}} \approx \ket{n}\ket{s_1}\ket{s_2}$. The expression in \ref{eq:time dependent identity} then becomes
\begin{equation}
\begin{split}
     \left(e^{iH_U t} \sigma_1 \cdot \sigma_2 e^{-iH_U t}\right)_{\Delta \omega = 0} = \sum_{n, n', s_1, s_1', s_2, s_2'} & \ket{n}\ket{s_1}\ket{s_2} \bra{n'}\bra{s_1'}\bra{s_2'} \bra{n_1}\bra{s_1}\bra{s_2} \sigma_1 \cdot \sigma_2 \ket{n'}\ket{s_1'}\ket{s_2'} \\
     &\times \delta_{n, n'} \delta_{s_1 + s_2, s_1' + s_2'}.
\end{split}
\end{equation}
It turns out that $\bra{s_1} \bra{s_2} \bra{n} \sigma_1 \cdot \sigma_2 \ket{n'} \ket{s_1'} \ket{s_2'} = 0$ for $n \neq n'$, and for $s_1 + s_2 \neq s_1' + s_2'$. Therefore,
\begin{equation}
    \left(U_1 U_2 e^{iH t}  \sigma_1 \cdot \sigma_2 e^{-iHt} U_2^\dagger U_1^\dagger \right)_{\Delta \omega = 0} = \sigma_1 \cdot \sigma_2 .
\end{equation}

\section{Calculation of Variance}
If a random variable $X$ can be written as a Taylor expansion in some small parameter $\lambda$, i.e.
\begin{equation}
    X = \sum_{n=1}^{\infty} \lambda^n Y_n .
\end{equation}
where $Y_n$ are random variables, then to second order in $\lambda$ the variance of $X$ is simply $\Var(X) = \lambda^2 \Var(Y)$. Therefore, to compute the variance of $T^\dagger A T$ we need only compute the variance of the terms of $T^\dagger A T$ which are linear in $\delta B(t)$. We get
\begin{align}
    \Var_{cl}(T^\dagger A T) &= \Var_{cl}\left(i\int_0^t dt' V_I(t')A - A V_I(t') \right), \\
    &= \Var_{cl}\left(i \int_0^t dt' (Q e^{-i \omega t} + Q^\dagger e^{i \omega t})A - A(Q e^{-i \omega t} + Q^\dagger e^{i \omega t})\right) \delta B(t'), \\
    &= \Var_{cl}(i ([Q, A] u + [Q^\dagger, A] u^*)). \\
\end{align}
Using the fact that $\expval{u}_{\delta B} = 0$, we re-write this expression as
\begin{align}
     \Var_{cl}(T^\dagger A T) &= \expval{\abs{\expval{[Q, A]u + [Q^\dagger, A] u^*}}^2}_{\delta B}, \\
     &= \expval{\left(\expval{[Q, A]}\expval{[A^\dagger, Q^\dagger]} + \expval{[Q^\dagger, A]}, \expval{[A^\dagger, Q]} \right) uu^*}_{\delta B}, \\
     &= 2 \left(\expval{[Q, A]}\expval{[A^\dagger, Q^\dagger]} + \expval{[Q^\dagger, A]}, \expval{[A^\dagger, Q]} \right) \Re(v) .
\end{align}
If $A$ is Hermitian, then this can be further simplified to
\begin{equation}
    \Var_{cl}(T^\dagger A T) = 4\expval{[Q^\dagger, A]}\expval{[A, Q]} \Re(v) .
\end{equation}

\section{Signal Variance Calculation}
For $Q_{\omega_0'} = \frac{\gamma_1}{2}\sigma_{-,1} + \frac{\gamma_2}{2}\sigma_{-,2}$, we need to calculate
\begin{equation}
    [\vec{\sigma_1} \cdot \vec{\sigma_2}, \sigma_{-, 1}] = \sigma_{z1} \sigma_{x2} - \sigma_{x1} \sigma_{z2} + i (\sigma_{y1} \sigma_{z2} - \sigma_{z1} \sigma_{y2}) .
\end{equation}
Then
\begin{equation}
    [\vec{\sigma_1} \cdot \vec{\sigma_2}, \frac{\gamma_1}{2}\sigma_{-,1} + \frac{\gamma_2}{2}\sigma_{-,2}] = \frac{1}{2} (\gamma_1 - \gamma_2) ( \sigma_{z1} \sigma_{x2} - \sigma_{x1} \sigma_{z2} + i (\sigma_{y1} \sigma_{z2} - \sigma_{z1} \sigma_{y2}) ) .
\end{equation}
Since $\vec{\sigma_1} \cdot \vec{\sigma_2}$ is Hermitian, we compute the magnitude squared of the above quantity to find $\Var_{cl}(T^\dagger \vec{\sigma_1} \cdot \vec{\sigma_2} T)$:
\begin{align}
    \Var_{cl}(T^\dagger \vec{\sigma_1} \cdot \vec{\sigma_2} T)_{\omega_0'} &= 4\expval{[Q^\dagger_{\omega_0'}, A]}\expval{[A, Q_{\omega_0'}]} \Re(v), \\
    &= \frac{1}{2} (\gamma_1 - \gamma_2)^2 \left(\expval{\sigma_{z1} \sigma_{x2} - \sigma_{x1} \sigma_{z2}}^2 + \expval{\sigma_{y1} \sigma_{z2} - \sigma_{z1} \sigma_{y2}}^2\right) S(\omega_0') t,\\
    &= \frac{1}{2} (\gamma_1 - \gamma_2)^2 \abs{\hat{z} \times \{\expval{\vec{\sigma_1}} \times \expval{\vec{\sigma_2}}\}}^2 S(\omega_0') t .
\end{align}

For $Q_{\omega} = \frac{\gamma_1 J_1(x_1) \omega_1}{2\omega}\sigma_{z1} + \frac{\gamma_2 J_1(x_2) \omega_2}{2\omega}\sigma_{z2}$, we get
\begin{equation}
    [\vec{\sigma_1} \cdot \vec{\sigma_2}, \sigma_{z1}] = 2 i (\sigma_{x1} \sigma_{y2} - \sigma_{y1} \sigma_{x2}) .
\end{equation}
Then
\begin{equation}
    \left[\vec{\sigma_1} \cdot \vec{\sigma_2}, \frac{\gamma_1 J_1(x_1) \omega_1}{2\omega}\sigma_{z1} + \frac{\gamma_2 J_1(x_2) \omega_2}{2\omega}\sigma_{z2} \right] = i \left(\frac{\gamma_1 J_1(x_1)\omega_1 - \gamma_2 J_1(x_2)\omega_2}{\omega} \right) (\sigma_{x1} \sigma_{y2} - \sigma_{y1} \sigma_{x2}) .
\end{equation}
Thus 
\begin{align}
    \Var_{cl}(T^\dagger \vec{\sigma_1} \cdot \vec{\sigma_2} T)_\omega &= 4\expval{[Q^\dagger_\omega, A]}\expval{[A, Q_\omega]} \Re(v), \\
    &= 2 \left(\frac{\gamma_1 J_1(x_1)\omega_1 - \gamma_2 J_1(x_2)\omega_2}{\omega} \right)^2  \expval{\sigma_{x1} \sigma_{y2} - \sigma_{y1} \sigma_{x2}}^2 S(\omega) t,\\
    &= 2 \left(\frac{\gamma_1 J_1(x_1)\omega_1 - \gamma_2 J_1(x_2)\omega_2}{\omega} \right)^2 \abs{\hat{z} \cdot \{\expval{\vec{\sigma_1}} \times \expval{\vec{\sigma_2}}\}}^2 S(\omega) t.
\end{align}

\section{Initial State of Magnetic Field \label{sec:dressingPhase}}
In this analysis, we assume that the magnetic field oscillates sinusoidally with amplitude $B_1$. However, this leaves some freedom in choosing the phase of magnetic field at time $t=0$. A magnetic field $B(t) = B_1 cos(\omega t - \phi)$ corresponds to the coherent state $\ket{\sqrt{\lambda} e^{i \phi}}$ where $\lambda$ is the average photon number. Because calculations in this work take place in the displaced basis (which is obtained by applying the unitary transformation $U$ to operators in the lab frame), we must also apply $U$ to the initial state $\ket{\alpha}\ket{s}$. We first show that for $\abs{\eta} \ll 1$ and $\abs{\alpha} \gg 1$,
\begin{equation}
    D(\eta)\ket{\alpha} \approx e^{\eta\alpha^* - \eta^*\alpha} \ket{\alpha} .
\end{equation}
We use the identity that for any complex numbers $\alpha$ and $\eta$,
\begin{equation}
    D(\eta)D(\alpha) = e^{\eta\alpha^* - \eta^*\alpha} D(\alpha) D(\eta) .
\end{equation}
Therefore,
\begin{align}
    D(\eta)\ket{\alpha} &= D(\eta)D(\alpha)\ket{0}, \\
    &= e^{\eta\alpha^* - \eta^*\alpha} D(\alpha) D(\eta) \ket{0}, \\
    &=  e^{\eta\alpha^* - \eta^*\alpha} D(\alpha) \ket{\eta}.
\end{align}
Coherent states are given explicitly in terms of the Fock states by
\begin{equation}
    \ket{\eta} = e^{-\abs{\eta}^2/2} \sum_{k=0}^{\infty} \frac{\eta^k}{\sqrt{k!}} \ket{k} .
\end{equation}
For $\eta \ll 1$ all but the $k=0$ terms are small, and so we make the approximation $\ket{\eta} \approx \ket{0}$. From this we conclude that
\begin{align}
    D(\eta)\ket{\alpha} &\approx e^{\eta\alpha^* - \eta^*\alpha} D(\alpha) \ket{0}, \\
    &\approx e^{\eta\alpha^* - \eta^*\alpha} \ket{\alpha} .
\end{align}
Now we examine what happens when we apply 
\begin{equation}
    U = D(\eta) \ket{+_x}\bra{+_x} + D(\eta)^\dagger \ket{-_x}\bra{-_x} ,
\end{equation}
where $\eta = \Omega/2\omega$ to the state $\ket{\alpha}\ket{s}$. We get
\begin{align}
    U\ket{\alpha}\ket{s} &= D(\eta) \ket{\alpha} \ket{+_x}\bra{+_x} \ket{s} + D(\eta)^\dagger \ket{\alpha} \ket{-_x}\bra{-_x} \ket{s}, \\
    &= \ket{\alpha} \left(e^{i\theta} \ket{+_x}\bra{+_x} \ket{s} + e^{-i\theta} \ket{-_x}\bra{-_x} \ket{s} \right), \\
    &= \ket{\alpha} \left(e^{i\theta} \ket{+_x}\bra{+_x} + e^{-i\theta} \ket{-_x}\bra{-_x} \right) \ket{s}, \\
    &= \ket{\alpha} e^{i\theta \sigma_x} \ket{s}, 
\end{align}
where $\theta = \Im(2 \eta^* \alpha)$. We thus see that an initial phase $\phi$ corresponds to a rotation by an angle $\theta$ about the $x$ axis.

%%RCD

\section{Robust dressing relaxation due to a relatively strong gradient field. \label{sec:robustRelaxation}} 
The dynamics of the system can be described by
\begin{align}
\dot{U}_I=e^{i\varphi_x\sigma_x/2}\mathbf{\omega_0\cdot\sigma}e^{-i\varphi_x\sigma_x/2}U_I ,
\end{align}
where 
\begin{align}
    \varphi  _x=\gamma_i\int_0^tB_1(t')dt' .
\end{align}

Within this interaction picture we can write longitudinal relaxation as the decay rate of $\expval{\sigma_x}$, which is given by time-dependent perturbation theory as
\begin{equation}
    \expval{\sigma_x} = 1 - \gamma^2 \Re{\int_0^t dt' \int_0^{t'} dt'' e^{-i \varphi_x(t')} e^{i \varphi_x(t'')} \expval{\delta B(t') \delta B(t'')}},
\end{equation}
where $\delta B(t)$ in this case represents fluctuations in either $B_z$ or $B_y$. In writing this approximation, we have taken the holding field to be small, which allows us to neglect the contribution to $T_1$ relaxation from $B_x$. The integral can be evaluated in terms of the Fourier transform of $e^{-i\varphi_x(t)}$, given by
\begin{equation}
    e^{-i\varphi_x(t)} = \sum_{n=-\infty}^{\infty} a_n e^{i n \omega_m t},
\end{equation}
where $\omega_m$ is the modulation frequency. By the same reasoning as in section \ref{sec:tdpt}, we need only consider terms that are the same frequency. Thus we get
\begin{align}
    \expval{\sigma_x} &= 1 - \gamma^2 \Re{\int_0^t dt' \int_0^{t'} dt'' \sum_n \abs{a_n}^2 e^{i n \omega_m (t' - t'')} \expval{\delta B(t') \delta B(t'')}}, \\
    &= 1 - 2t \sum_n \abs{a_n}^2 S(n \omega_m) .
\end{align}
Therefore we get
\begin{align}
    \frac{1}{T_1} &= \frac{\gamma^2}{2} \sum_n \abs{a_n}^2 S(n \omega_m), \\
     &\approx \gamma^2 (G_z^2 + G_y^2) D \sum_n \frac{\abs{a_n}^2}{(\omega_m n)^2},
\end{align}
where in the last step we took the diffusion approximation to evaluate $S(n \omega_m)$. If the system is not in the diffusion limit a more exact formulation can be implemented, for example, the spectrum of the correlation function presented in reference \cite{swank2016} would allow accurate predictions from the ballistic through the diffusive regimes. 

\bibliography{references}

\end{document}